\documentclass[useAMS,usenatbib]{mn2e}

\usepackage{mathptmx}
\usepackage[hypertex]{hyperref}

\usepackage{graphicx}

\newcommand{\rband}{{\it r\/}-band}
\newcommand{\zband}{{\it z\/}-band}

\newcommand{\unit}[1]{\mathrm{#1}}
\newcommand{\unitp}[2]{\ensuremath{\mathrm{#1}^{#2}}}

\title[Environment \& early-type galaxies]{The star formation history
of early-type galaxies as a function of mass and environment}

\author[Clemens et al.]{M.S. Clemens$^{1}$, A. Bressan$^{1,2}$, B. Nikolic$^{3}$, P. Alexander$^{4}$, F. Annibali$^{2}$, R. Rampazzo$^{1}$\\
$^{1}$INAF-Osservatorio Astronomico di Padova, Vicolo dell'Osservatorio, 5, 35122 Padova, Italy.\\
$^{2}$SISSA-ISAS, International School for Advanced Studies, ia Beirut 4, 34014 Trieste, Italy\\
$^{3}$National Radio Astronomy Observatory, Green Bank, WV (GB) and Charlottesville, VA (CV), USA\\
$^{4}$Astrophysics Group, Cavendish Laboratory, Madingley Road, Cambridge CB3 0HE, UK}
\date{2005-03-04}
\pagerange{\pageref{firstpage}--\pageref{lastpage}; } \pubyear{2005}

\begin{document}

\maketitle

\label{firstpage}

\begin{abstract}
Using the third data release of the Sloan Digital Sky Survey (SDSS) we have rigorously
defined a volume limited sample of early-type galaxies in the redshift range
$0.005<z\leq 0.1$. We have defined the density of the local environment for each
galaxy using a method which takes account of the redshift bias introduced by
survey boundaries if traditional methods are used. At luminosities greater than
our absolute r-band magnitude cutoff of -20.45 the mean density of environment
shows no trend with redshift. We calculate the Lick indices for the entire sample
and correct for aperture effects and velocity dispersion in a model independent
way. Although we find no dependence of redshift or luminosity with environment we
do find that the mean velocity dispersion, $\sigma$, of early-type galaxies in
dense environments tends to be higher than in low density environments. Taking
account of this effect we find that several indices show small but very
significant trends with environment that are not the result of the correlation between indices and
velocity dispersion. 
The statistical significance of the data is sufficiently high to reveal
that models accounting only for $\alpha$-enhancement
struggle to produce a consistent picture of age and metallicity of the sample galaxies,
whereas a model that also includes carbon enhancement fares much better. 
We find that early-type galaxies in the field
are younger than those in environments typical of clusters but that neither metallicity, $\alpha$-enhancement nor carbon enhancement are influenced by the environment. The youngest early-type galaxies in both field and cluster environments are those with the lowest $\sigma$. However, there is some evidence that the objects with the largest $\sigma$ are slightly younger, especially in denser environments.

Independent of environment both the metallicity and
$\alpha$-enhancement grow monotonically with $\sigma$. This suggests
that the typical length of a the star formation episodes which formed
the stars of early-type galaxies decreases with $\sigma$. More massive
galaxies were formed in faster bursts.
 
We argue that the timing of the process of formation of early-type
galaxies is determined by the environment while the details of the
process of star formation which has built up the stellar mass are
entirely regulated by the halo mass. These results suggest that the
star formation took place \emph{after} the mass assembly and favours
an anti-hierarchical model. In such a model the majority of the
mergers must take place before the bulk of the stars form. This can
only happen if there exists an efficient feedback mechanism which
inhibits the star formation in low mass halos and is progressively
reduced as mergers increase the mass.

\end{abstract}


\section{Introduction}

Present-day spheroids are thought to contain 
a large fraction, $\simeq$ 60\%-80\%, 
of the total stellar mass in the Universe 
(Fukugita \& Peebles 2004).
They are thus fundamental tracers of the assembly
of baryonic matter in the different environments
generated from primordial density fluctuations.

There are well established
scaling relations indicating that the 
gravitational potential well is one of the key drivers
in the formation and evolution of these galaxies. 
Suffice here to recall
the colour-magnitude relation of early-type
galaxies in clusters (Bower, Lucey \& Ellis  1992),
the relation between colour and central velocity dispersion
 and the fundamental plane (Bender, Burstein \& Faber 1993).
In contrast, the role played by the environment is not yet clear

The environment may affect galaxy properties in two distinct ways:
first, the initial conditions from which
galaxies form are different and second their
subsequent evolution may be affected by processes whose efficiency
depends on environment.

As for the initial conditions, the hierarchical 
clustering paradigm predicts
a very simple dependence on environment, 
in the sense that objects in denser
regions form earlier (Lacey \& Cole 1994). 
Subsequently the environment (whether we mean the local galaxy density
or membership of a cluster) may affect galaxy properties in several ways.

In galaxy clusters, the large scale gravitational potential 
is directly responsible for so-called `galaxy
harassment' which describes the cumulative effects of high
speed encounters on the dynamics and star formation rate
of the galaxies (Moore et al. 1996). 
Furthermore, the intra-cluster medium can 
remove high filling factor ISM
from disk galaxies (Gunn \& Gott 1972). 
Initial conditions and subsequent environmental effects  
are the most likely
explanation for the existence of the colour magnitude relation
(Bower, Lucey and Ellis 1992), and the morphology density relation
(Dressler, 1980, McIntosh, Rix \& Caldwell 2004; Balogh et al. 2004; Hogg
et al. 2004; Croton et al. 2005), respectively.

In the field however, the separation of the  different aspects of the
environmental influence is rendered much less clear by the
observational difficulty of obtaining homogeneous samples.  Various
authors, using samples of $\sim 100$ galaxies, find evidence that
early-type galaxies in low density environments are younger and more
metal rich than those in clusters (Longhetti et al. 2000, Thomas et
al. 2005, Denicol\'o et al. 2005). Severeal authors (E.g. Kuntschner et al 2002, Denicol\'o
et al. 2005, Thomas et al. 2005) also find that
more massive and older galaxies have increased $[\alpha/{\rm Fe}]$
values, suggesting more rapid star formation
time-scales. S\'anchez-Bl\'azquez et al. (2003) use a sample of 98
galaxies to find lower C4668 and CN2 index values in the centre of the
Coma cluster than in the field.

However, large and homogeneous galaxy samples are needed to
provide the statistical discrimination necessary to identify small
changes among stellar populations.
Surveys such as the 2dF Galaxy Redshift Survey, Las Campanas Redshift
Survey and the ongoing Sloan Digital Sky Survey (SDSS) have all been used
for such studies.
A sample
of 1823 galaxies (of all types) has been used by Balogh et al. (1999) to
find stronger 4000 Angstrom break, and weaker [OII] and $\rm H\delta$
indices in the cluster environment. These authors conclude that the last
star formation episode occurred more recently in low density environments.
Bernardi et al. (1998), with a sample 931 early-type galaxies found only a
minimal difference in the Mg2$-\sigma$ relation, implying that field
galaxies are only approximately 1 Gyr younger than those in clusters.

Bernardi et al. (2005a) have used a sample of over 39,000 galaxies from the SDSS to find that age correlates with velocity dispersion from an analysis of the colour-magnitude-velocity dispersion relation. Based on a similar sample Bernardi et al. (2006) analyse Lick indices to conclude that early-type galaxies in dense environments are less than 1 Gyr older than field objects and have a similar metallicity. Similar results have also been obtained by Smith et al. (2006) and Nelan et al. (2005).

In this paper we study the absorption line indices of a large sample of
early-type galaxies selected from the third release of the SDSS.
We couple the high statistical significance of the sample with the most
recent stellar population models to obtain an unbiased picture of the
history of star formation in spheroids, as a function of their velocity dispersion,
in different environments.

The paper is organised as follows. Section~2 is devoted to
the sample selection and definition of environment. In Section~3
we derive the narrow band indices and describe the various
corrections to transform them to the Lick IDS system.
In Section~4 we analyse the data by means of our new simple stellar population 
models. Our findings are discussed in Section~5. Finally, 
in the conclusion Section we present an evolutionary scenario
as a function of galaxy mass and environment.

\section{The sample and data}

\subsection{Sample selection}

The primary sample used in this study was selected along similar lines
to the sample used by G\'omez et al. (2003) and
Nikolic, Cullen \& Alexander (2004), i.e., it is a luminosity- and (pseudo)
volume-limited sample. The parent sample consisted of all objects, as
of Data Release 3 (DR3), observed spectroscopically by the SDSS and
classified as galaxies\footnote{These are all of the objects in files
gals-DR2.fit.gz and gals-DR3.fit.gz available at
\href{http://das.sdss.org/DR3/data/spectro/ss_tar_23/}{http://das.sdss.org/DR3/data/spectro/ss\_tar\_23/}.}.
In practice, all of the galaxies that are part of our sample, given
its selection criteria, should have been selected for spectroscopic
observations as a part of the of SDSS Main Galaxy Sample (MGS)
described by Strauss et al. (2002). Before selecting the
volume-limited sample a small number of spurious entries were
identified and removed: galaxies within 5 arc seconds of other galaxies
(most likely de-blending errors), galaxies with redshift confidence of
less than 0.7 (they cannot be reliably placed in the volume limited
sample) and galaxies with \zband\ magnitudes fainter than 22.8, i.e.,
not reliably detected in the \zband\ (these are most likely spurious
detections).

The volume limited sample was then constructed by selecting galaxies
in the redshift range $0.005<z \leq 0.1$ and with Petrosian absolute
magnitudes brighter than -20.45. Since the \rband\ magnitude limit of
the MGS is 17.7, the absolute magnitude limit adopted should avoid a
Malmquist bias in the volume limited sample.

In order to limit our sample to include only early-type galaxies we
then adopted the following selection procedure:  1) The compactness
ratio as defined in the SDSS was constrained to be less than 0.33 (see
Shimasaku et al. 2001).  2) Any object with detected emission in the
H$\alpha$ line was excluded in order to minimise the effects of Balmer
emission line infilling.  This is particularly important for the
correct interpretation of the H$\beta$ absorption feature.  To this
sample we further restricted the objects of study to, 3) those objects
with a well measured velocity dispersion, and 4) those objects where
the density of the environment could be well determined, independently
of survey boundaries (see below). After such selection our sample
consisted of 3614 galaxies. In section~\ref{sect:regression} we
describe a further selection based on the signal-to-noise ratio of the
specttra to be included in the final regression analysis.

\subsection{Estimating density of environment}

The algorithm used to quantify the density of the environment is
similar to that used by Dressler (1980) in his original
study of the relationship between density and morphology: it is based
on finding the projected separation to the $n$-th nearest neighbour,
$r_{n}$. This distance can, if required, be used to estimate the
surface density of galaxies via $\sigma_{{\rm D},n} = n / r_{n}^{2} \pi$.

Unlike the original algorithm, the one presently employed makes use of
the measured redshifts of galaxies to search for nearest neighbours
only in the redshift shell $\Delta v= \pm 1200\,\unit{km}\,\unitp{s}{-1}$ 
around the galaxy being considered following G\'omez et al. (2003); Goto et al. (2003) 
(although our maximum velocity difference is slightly larger). Because of this
modification, no correction for the background galaxy field density
need be made. Since in our approach redshift information is
required, the search for $n$-th nearest neighbour is necessarily
restricted to a sample of galaxies with measured redshifts. To avoid a
dependence of the calculated density scale on redshift, we search for
$n$-th nearest neighbour in the volume-limited catalogue itself. A
photometric approach to calculating galaxy density in the SDSS is
discussed by Hogg et al. (2003).

Since density of environment is a non-local measure, survey boundaries
will inevitably present a problem.  Galaxies close to one of the
boundaries will have the separation to their $n$-th neighbour (the
density metric that we are using in this study) overestimated and
therefore their density of environment underestimated; this effect is
quite clearly illustrated in the simulations presented by Miller et al.
 (2003). Hence galaxies close to survey edges have
unreliably measured densities and should not be considered in further
analysis.  The non-compact nature of the SDSS survey area makes this
effect quite pronounced, although less so in DR3 than in earlier releases. 

\begin{figure}
\includegraphics[clip,width=0.95\columnwidth]{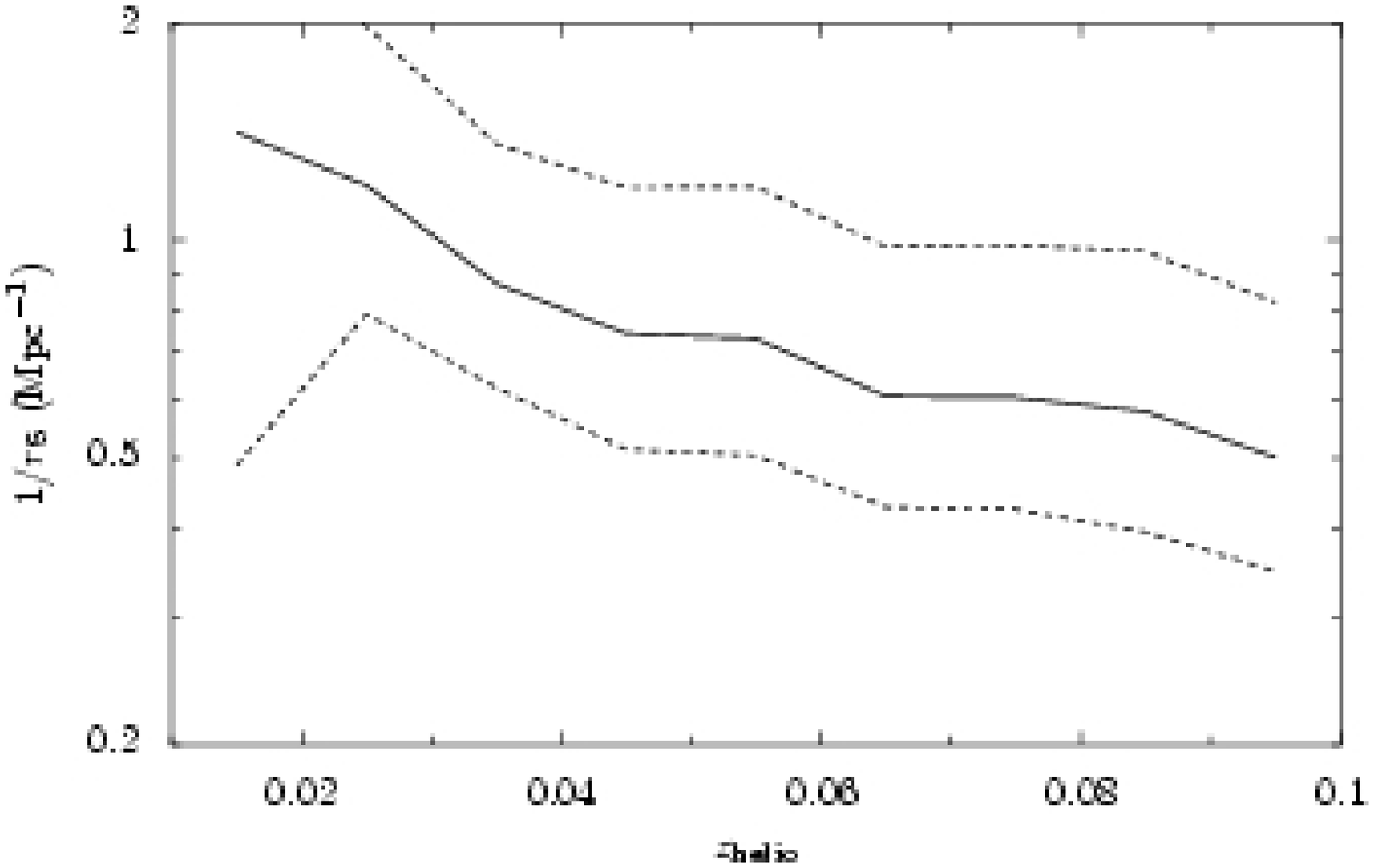}
\includegraphics[clip,width=0.95\columnwidth]{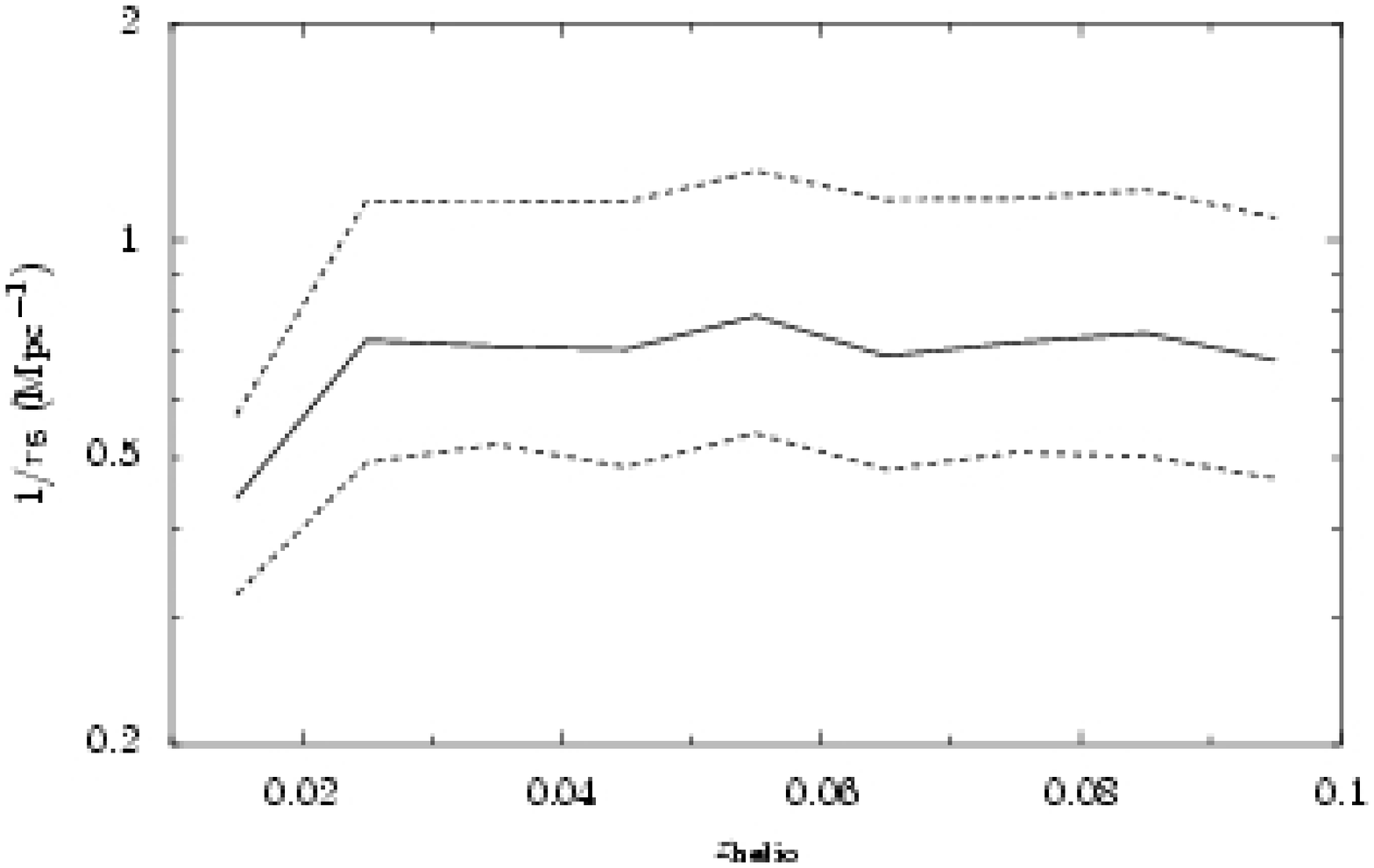}
\caption{The distributions---represented as the median (solid line)
  and the 25 and 75 percent quartiles (dashed lines)---of density as a
  function of redshift for samples corrected for the effects of survey
  boundaries in the traditional way (top) and using the technique
  described in the text (bottom).}
\label{fig:densrs}
\end{figure}

The usual approach to dealing with this problem is to remove from
further analysis galaxies with $n$-th neighbours at a larger separation
than a survey boundary.  This was shown by Miller et al.
 (2003) to be effective, but such an approach also preferentially removes
galaxies in low density environments at low redshift compared to high
redshift, as shown in the upper panel of Fig.~\ref{fig:densrs}.
This can be understood by considering a population of
galaxies with a fixed physical distance to their $n$-th neighbour:
with increasing redshift, the \emph{angular} separation to a survey
boundary can be smaller and still allow good estimation of density;
therefore the necessary margin around the survey boundary decreases
and the effective useful area of the survey increases with increasing
redshift.

The correlation between density and redshift in the sample after such
an edge-correction procedure can be avoided by removing a somewhat
larger number of galaxies as follows. If $r_{n}$ is the separation to
the n-th neighbour and $r_{\rm E}$ is the separation to a survey edge,
only galaxies which satisfy $r_{n} \frac{z}{z_{0}} \leq r_{\rm E}$
should be retained in the edge-corrected sample, where $z_{0}$ is a
fiducial redshift which should also be (close to) the lower redshift
limit of the sample. 

Finally, the choice of $n$ to use as a density metric will also
determine how much of an effect survey boundaries have. We follow
Goto et al.(2003) and use $n=5$ throughout. 
The edge-corrected sample is shown in the lower
panel of Fig.~\ref{fig:densrs}.

\begin{figure}
\includegraphics[clip,width=8.2cm]{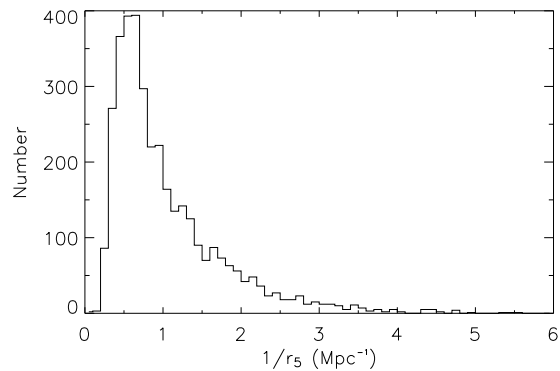}
\caption{The distribution of the environmental density for our sample.}
\label{fig:PDens}
\end{figure}

Fig.~\ref{fig:PDens} shows the distribution of our measure of environment for all
the galaxies in our sample.

\begin{figure}
\includegraphics[clip,width=0.95\columnwidth]{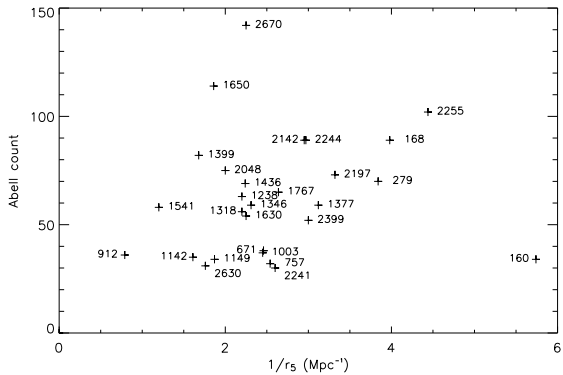}
\caption{Comparison of our measure of environment, $1/r_5$, with the number of cluster members as given in Abell. The values of $1/r_5$ are the mean value for the galaxies in our sample associated with each cluster. The Abell cluster designation is shown near each point.}
\label{fig:clusters}
\end{figure}

In Fig.~\ref{fig:clusters} we compare our measure of the local galaxy density
with the number of cluster members for those Abell clusters where our sample
contains at least one galaxy. There is a correlation between the measures but one
is not a good predictor of the other, and neither would one especially expect
them to be. Our measure of environment quantifies the \emph{local} environment of
a galaxy whereas the total number of galaxies in an Abell cluster is a measure of
the large-scale environment (recall also that we define environment using only
those galaxies bright enough to be detected at all redshifts in our sample). The
most important point about Fig.~\ref{fig:clusters} is that it shows that galaxies
in our sample that have a density parameter above 1 are in environments typical
of galaxy clusters. 38\% of our galaxies have a value of $1/r_5>1.0$, and 10\%
$1/r_5>2.0$.

As a further check for the reliability of our measure of environment we visually
checked the fields of our sample galaxies with the 20 highest and 20 lowest
values of $1/r_5$. This confirmed that low values do indeed represent
galaxies in poor environments and visa-versa.

\section{Calculating the line indices}

Since one of the final goals of this study is to interpret the indices
with models based on the Lick system we first transform the SDSS
spectra to the wavelength dependent resolution of the Lick-IDS
spectra. This corresponds  to $\rm \sim 11.5\;\AA$ at $\rm 4000\;\AA$,
$\rm \sim 8.4\;\AA$ at $\rm 5000\;\AA$ and $\rm \sim 9.8\;\AA$ at $\rm
6000\;\AA$.

From the SDSS spectra we then extract the 21 line--strength indices of
the original Lick-IDS system (Trager et al. 1998)  plus the additional
indices H$\gamma$A, H$\delta$A and H$\gamma$F,  H$\delta$F (Worthey \&
Ottaviani 1997) and the indices B4000 (Hamilton, 1985) and HK (Rose, 1984, 1985).  Our extraction  makes
use of the redshift of each object as provided in the SDSS.  For this
extended Lick system we transform the passband definition of Trager et
al.  (1998) to the vacuum (SDSS spectra are given in terms of vacuum
wavelengths).  A comparison with the indices in common with the
associated SDSS tables show that they lie on a one-to-one relation.

In our analysis the transformation to the Lick system, which is
usually made by comparing selected observed standard stars with the
Lick IDS observations, is not possible. This is simply because of the
lack of observations of standard Lick stars in the adopted SDSS data
release.  For this reason we work with index differences,  i.e. we do
not consider absolute values but only trends  with respect to a
reference population. This strategy  avoids significant biases in our
analysis, because it has been shown that  the transformations to the
Lick IDS standard stars  mostly consist of an offset correction (see
e.g. Puzia et al. 2002,  Rampazzo et al. 2005). Our results will also
be much less dependent on the zero point calibration of the SSP models.

The errors on each index measurement were calculated by first
measuring the rms noise in each spectrum and then reproducing a model
spectrum with the same rms. We then used a Monte-Carlo method to
measure the indices for 200 realisations of the spectrum. We define
the error on our measurement of an index as the standard deviation of
the index values so obtained.

\subsection{Velocity dispersion correction}

Because we are interested in the intrinsic index-$\sigma$ relation for early-type
galaxies it is important that no bias be introduced by an inappropriate velocity
dispersion correction. However, as we show in the appendix, although most indices show some
correlation with the velocity dispersion of a galaxy, and velocity dispersion
is in turn correlated with environment, our results on the \emph{influence of the
environment} are not sensitive to our velocity dispersion correction. The
velocity dispersion correction is thus important only for the shape of the index-$\sigma$ relations.
The velocity dispersion for each galaxy is provided in the SDSS.

To ensure that our derived index-$\sigma$ relations are not a result of an error
in velocity dispersion correction we have made the correction in two different
ways and carried out our analysis with both versions.

In the first we simply smooth all the spectra to a velocity resolution of
$400\;\rm km\,s^{-1}$. Although this will result in index values which are all
lower than they would be at say the resolution of the Lick system, we are sure that we have not
introduced any model dependent effects.

In the second method we have smoothed the spectra of G and K-type giant stars
observed by the SDSS to a number of different velocity dispersions
from zero to $400\;\rm km\,s^{-1}$. 
For each index of a given galaxy we identify a stellar spectrum 
with the same velocity dispersion and 
with similar line index values, and then use this star to make
the correction to the Lick IDS system.

A comparison between these 2 methods has been made by using the stars observed by
the SDSS to correct indices to a velocity dispersion of $400\;\rm km\,s^{-1}$ and
comparing to the results of the first method. Only in some of the bluer indices (Ca4227, G4300, Fe4383 and Ca4455) was any difference seen in the resulting index-$\sigma$ relation.

\subsection{Aperture correction}
Every spectra in the SDSS is taken through a fibre with a diameter of
$3^{\prime\prime}$ and so the physical radius sampled for any given galaxy is a
function of distance. We must apply an aperture correction to remove this effect.
Rampazzo et al. (2005) have measured the spatial index gradients in 50 elliptical galaxies and
we have used the mean radial gradients found for each index to correct our index values to 
the standard radius of $r_{\rm e}/10$ where $r_e$ is the effective radius. Using the data of Rampazzo et al. we take $(I(r_e/2) - I(r_e/10))/I(r_e/10)$ as a measure of the spatial index gradient and plot this against $\log \sigma$ for all 50 galaxies in their sample. A least-squares, straight line fit to these plots is then found;

\begin{equation}
\frac{(I(r_e/2)-I(r_e/10))}{I(r_e/10)} = x \log(\sigma/100) + y
\label{eq:ap_cor1}
\end{equation}

\noindent The values of $x$ and $y$ for each index are reported in table~\ref{tab:ap_cor_fits}. The SDSS gives values for the half light radius ($r_e$) of each galaxy and so we can correct the index measured with a $3^{\prime\prime}$ SDSS fibre to the standard radius of $(r_{\rm e}/10)$ by assuming a linear change in index value with radius;

\begin{equation}
I(r_e/10) = \frac{I({\rm SDSS})}{\big(x \log(\sigma/100) + y \big) \big( \frac{\log (15/r_e)}{\log 5}\big)\; +1}
\label{eq:ap_cor2}
\end{equation}

\noindent where $I({\rm SDSS})$, the index measured from the SDSS spectrum, is the index measured within a radius of $1\farcs 5$. $\sigma$ is in units of $\rm km\,s^{-1}$ and $r_e$ in arcseconds.  

\begin{table}
\caption{Parameter values for fits to the radial index variations measured by Rampazzo et al. (2005). Values for $x$ and $y$ in equation~\ref{eq:ap_cor1} are given.}
\label{tab:ap_cor_fits}
\begin{tabular}{lrrlrr}
\hline
Index & x & y & Index & x & y \\
\hline
CN1       & 0.468   & -0.434  & Fe5270    & 0.195   & -0.0937 \\
CN2       & 0.658   & -0.463  & Fe5335    & 0.0340  & -0.0565 \\
Ca4227    & 0.441   & -0.0999 & Fe5406    & -0.0990 & -0.0204 \\
G4300     & 0.129   & -0.0531 & Fe5709    & 0.131   & -0.0225 \\
Fe4383    & -0.0344 & -0.0570 & Fe5782    & 0.227   & -0.0986 \\
Ca4455    & -0.111  & -0.0716 & NaD       & 0.127   & -0.193 \\
Fe4531    & 0.103   & -0.0478 & TiO1      & 0.143   & -0.108 \\
C4668     & 0.318   & -0.219  & TiO2      & 0.0434  & -0.0573 \\
H$\beta$  & 0.410   & -0.0369 & B4000     & -0.0785 & 0.0574 \\
Fe5015    & 0.517   & -0.158  & HK        & 0.0165  & 0.00396 \\
Mg1       & 0.132   & -0.142  & HDF       & 1.18    & -0.539 \\
Mg2       & 0.131   & -0.107  & HGF       & 0.126   & -0.183 \\
Mgb       & -0.0068 & -0.0594 & & & \\
\hline
\end{tabular}
\end{table}

\begin{figure*}
\includegraphics[clip,width=17cm]{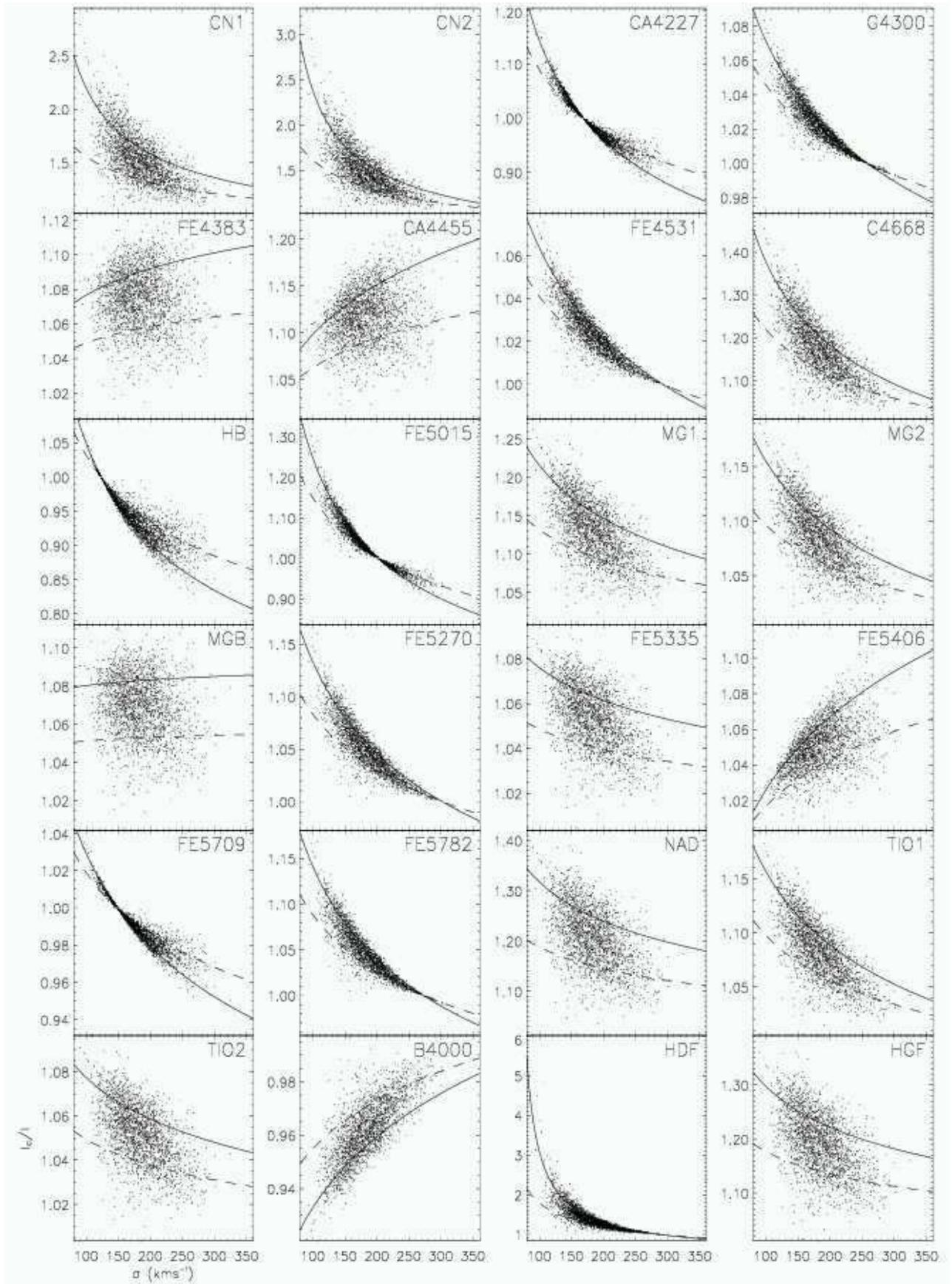}
\caption{Aperture corrections as a function of velocity dispersion, $\sigma$. The points are the data, and the solid and dashed lines show the correction applied for an effective radius of 2$^{\prime\prime}$ and 4$^{\prime\prime}$ respectively.}
\label{fig:ap_cor}
\end{figure*}

Fig.~\ref{fig:ap_cor} shows the aperture 
corrections applied for each index. The accuracy of the aperture corrections depends
on the accuracy of the value for the half light radius reported in the SDSS. This in
turn will depend principally on the seeing. The fractional error will therefore increase
with redshift, on average, as the angular sizes of the galaxies decreases, with the radii tending to be increasingly overestimated at greater distances. This effect has 
yet to be quantified in the current release of the SDSS and the effect would influence the gradients of our index-$\sigma$ plots (Fig.~\ref{fig:sig-index}). However, as we see below, our derived index-$\sigma$ relations agree very well with those derived from other studies based on more local samples and so we expect the influence of a seeing-induced effect to be small. Further, we note that because our 
various sub-samples have similar redshift distributions (Fig.~\ref{fig:histograms}) our results on the influence of environment are insensitive to the gradient of the index-$\sigma$ relations (see appendix).

\section{results}

Below we quantify our results making use of both simulations and models. We stress however that the results are qualitatively apparent in the data \emph{before the use of either}. In the log$\sigma$-index plane, for example, several indices show a similar gradient for different environments but with an offset in index value. This immediately implies an effect of environment on the index that is not due to the elevated $\sigma$ in denser environments (see section~\ref{enveffect}). Because we have taken great care to minimise selection bias we are able to treat the sample as a whole for the statistical analysis.     

The velocity dispersion, $\sigma$, does not have a one-to-one relation to the mass of the galaxy. Recent work by Cappellari et al. (2006) has investigated the relation between galaxy mass-to-light (M/L) ratio and the line-of-sight component of the velocity dispersion within the effective radius. They provide relations (their equations 7 and 10) which allow a transformation from measured velocity dispersion to galaxy mass;

\begin{equation}
M_{10} = (16.5 \pm 7.8)\;\sigma_{200}^{3.11 \pm 0.43}
\end{equation}

\noindent where $M_{10}$ is the galaxy mass in units of $10^{10}\,\rm M_{\odot}$ and $\sigma_{200}$ is theluminosity weighted mean velocity dispersion within one effective radius in units of $200 \;\rm km\,s^{-1}$. The errors have been propagated from the parameter values found by Cappellari et al. (2006). We do not make any conversion to mass but those who wish to can apply this formula.

\subsection{Trends of line indices with velocity dispersion}
\begin{figure*}
\includegraphics[clip,width=16cm]{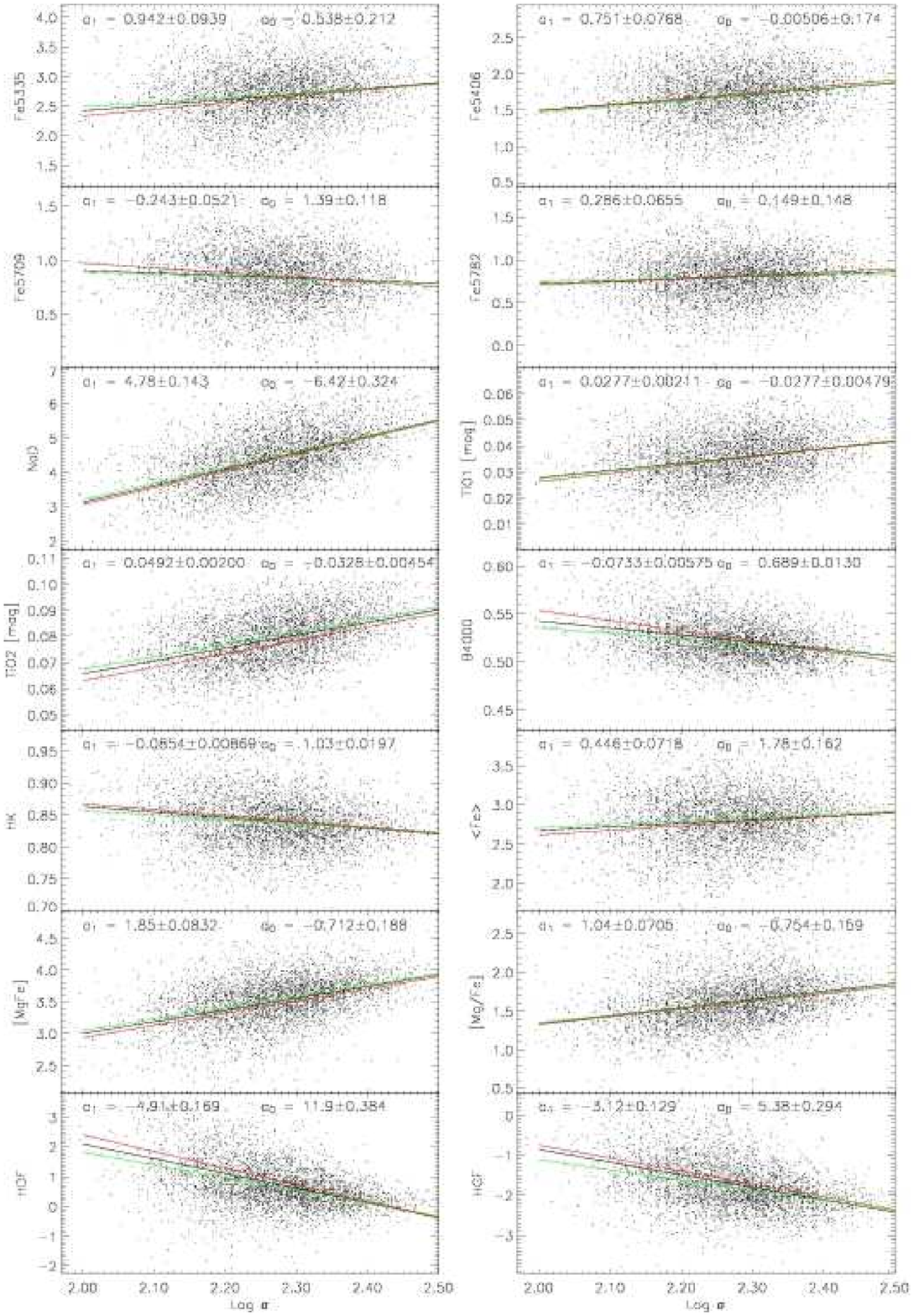}
\caption{Index-$\sigma$ relations. The data have been corrected for both aperture and velocity dispersion effects. The black line shows a least squares, straight line fit to the whole sample, the parameters of which are reported at the top of each panel. The red and green lines show fits to low and high density sub-samples with $1/r_5 < 0.5$ and $1/r_5 > 1.5$ respectively.}
\label{fig:sig-index}
\end{figure*}

Figure ~\ref{fig:sig-index} shows the index-$\sigma$ relations for our entire sample after the aperture and velocity dispersion corrections have been made. The gradients of the relations are reported for a simple first order fit along with the corresponding errors. Most indices show some correlation with the velocity dispersion.

The following indices show a significant ($>5\sigma$) positive correlation: CN1,
CN2, G4300, Fe4383, Ca4455, Fe4531, C4668, Mg1, Mg2, Mgb, Fe5335, Fe5406, NaD, TiO1,
TiO2, $<$Fe$>$, [MgFe] (Gonz\'alez, 1993) and [Mg/Fe] (defined as $Mgb/(0.5(Fe5270 + Fe5335))$). Negative gradients are instead seen for: $\rm H\beta$,
$\rm H\delta$, $\rm H\gamma$, Fe5015, B4000 
(note the definition of this index in the SDSS is inverted with respect to that of many authors) and HK. No appreciable gradient is evident for: Ca4227, Fe5270, Fe5709 and Fe5782. Below we consider straight line fits of the form $I = a_{0} + a_{1}\log \sigma$ or $\log I = A_{0} + A_{1}\log \sigma$.

Most work on the trends of indices with velocity dispersion has been done for the Mg2-$\sigma$ relation for which we find $a_1 = 0.209 \pm 0.0056$ and $a_0 = -0.198 \pm 0.013$. Bernardi et al. (1998) find $a_1 = 0.224 \pm 0.008$ and $a_0 = -0.23 \pm 0.019$ for a sample of 931 galaxies and Kuntschner et al. (2002) find corresponding values of $0.216 \pm 0.025$ and $-0.236 \pm 0.057$ for Fornax spheroidals. Both studies give fits consistent with ours. Jorgensen (1997) finds values of $a_1 = 0.196 \pm 0.009$ and $a_0 = -0.155$. The slope is in agreement with our value whereas the intercept seems to be slightly lower (although this author quotes no error on the intercept value). 

For MgB we find $a_1 = 3.78 \pm 0.153$ ($A_1 = 0.389 \pm 0.0165$) which agrees well with Bernardi et al. (2003) and is similar to the relation found by Thomas et al. (2004) (although they do not explicitly quote the gradient). 

For $<$Fe$>$ we find $a_1 = 0.449 \pm 0.072$ and $a_0 = 1.78 \pm 0.162$ ($A_1 = 0.0820 \pm 0.0118$ and $A_0 = 0.258  \pm  0.0267$) respectively. These values are both consistent with that found by Bernardi et al. (2003) and Jorgensen (1997).   

Our [Mg/Fe]-$\sigma$ relation has $a_1 = 1.04  \pm  0.0700$, $a_0 = -0.747 \pm 0.159$ ($A_1 = 0.306  \pm  0.0194$, $A_0 = -0.499 \pm 0.044$). Kuntschner et al. (2002) find values in the range $A_1 \sim 0.18 - 0.30$ for $< 100$ early-type galaxies in Virgo, Coma and Fornax. Our value is consistent with their findings despite the fact that our value refers to early-type galaxies in a wide range of environments. Our [Mg/Fe]-$\sigma$ relation is marginally consistent with Bernardi et al. (2003) who find $A_1 = 0.22 \pm 0.04$. Trager et al. (2000) find $A_1 = 0.33 \pm 0.01$ and $A_0 = -0.58 \pm 0.01$ which is similar to the gradient we find but with a slightly lower intercept.

We obtain a slope for the $\rm H\beta$ index of $a_1 = -1.75 \pm 0.0670$ ($A_1 = -0.450 \pm  0.0191$). This is similar to that found by Thomas et al. (2005) ($\sim -1.7$) but is somewhat steeper than that found by Bernardi et al. (2003), who obtain $d(\log \rm{H}\beta)/d(\log \sigma) = -0.24 \pm 0.03$.

A thorough treatment of the index-$\sigma$ relations in cluster galaxies has been carried out by Nelan et al. (2005) (their table 7). Most of the gradients measured by these authors are similar to the gradients we find for our \emph{high density} sub-sample, but some indices show large differences from our measured trends. These are Ca4227, C4668, Fe5015 and Fe5270. We note that apart from the differences in sample characteristics these authors apply an aperture correction to a fixed physical radius that is not a function of $\sigma$; our correction instead is to $r_{\rm e}/10$ and is applied as a function of velocity dispersion. Aperture corrections have not traditionally taken into account the fact that radial index gradients are a function of $\sigma$. We note that the indices in which we see the most significant disagreement tend to have aperture corrections that vary more strongly with $\sigma$ (Fig.~\ref{fig:ap_cor}).

In section~\ref{sect:regression}, these relationships are
interpreted in terms of physical parameters of the galaxies.

\begin{figure*}
\includegraphics[clip,width=16cm]{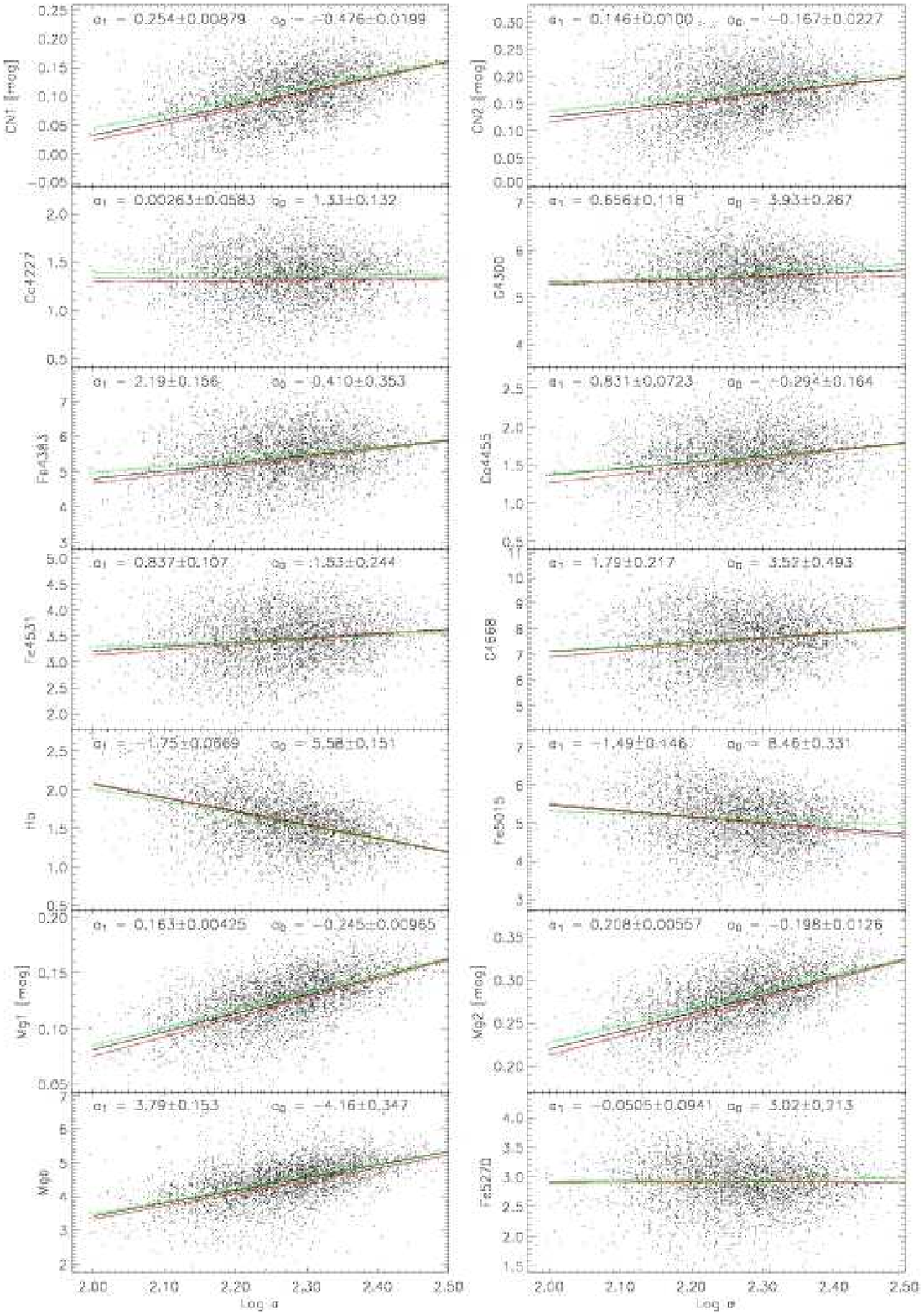}
\end{figure*}

\subsection{The effect of environment on the line indices} 
\label{enveffect}
Having corrected our catalogue for the various biases discussed above we then
divided it in terms of the density parameter $1/r_5$ to investigate the effect of
environment on the line indices. Of course, \emph{if the distribution of our sub-samples 
differs in any parameter that is correlated with the index we must take
account of this}.

We divided our total sample of 3614 galaxies in the following way. First we
selected a low density sample made up of those galaxies with $1/r_5 < 0.5$ (726
objects) against which to compare galaxies in increasingly dense environments.
These were then defined as follows: 1) $0.5<1/r_5<1.0$ (1513 objects), 2)
$1/r_5>0.5$ (2888 objects), 3) $1/r_5>1.0$ (1362 objects), 4) $1/r_5>1.5$ (706
objects), and 5) $1/r_5>2.0$ (357 objects). The upper density bound on the first
of these was intended to search for effects at environmental densities below
those typical of clusters.

\begin{figure*}
\includegraphics[clip,width=18cm]{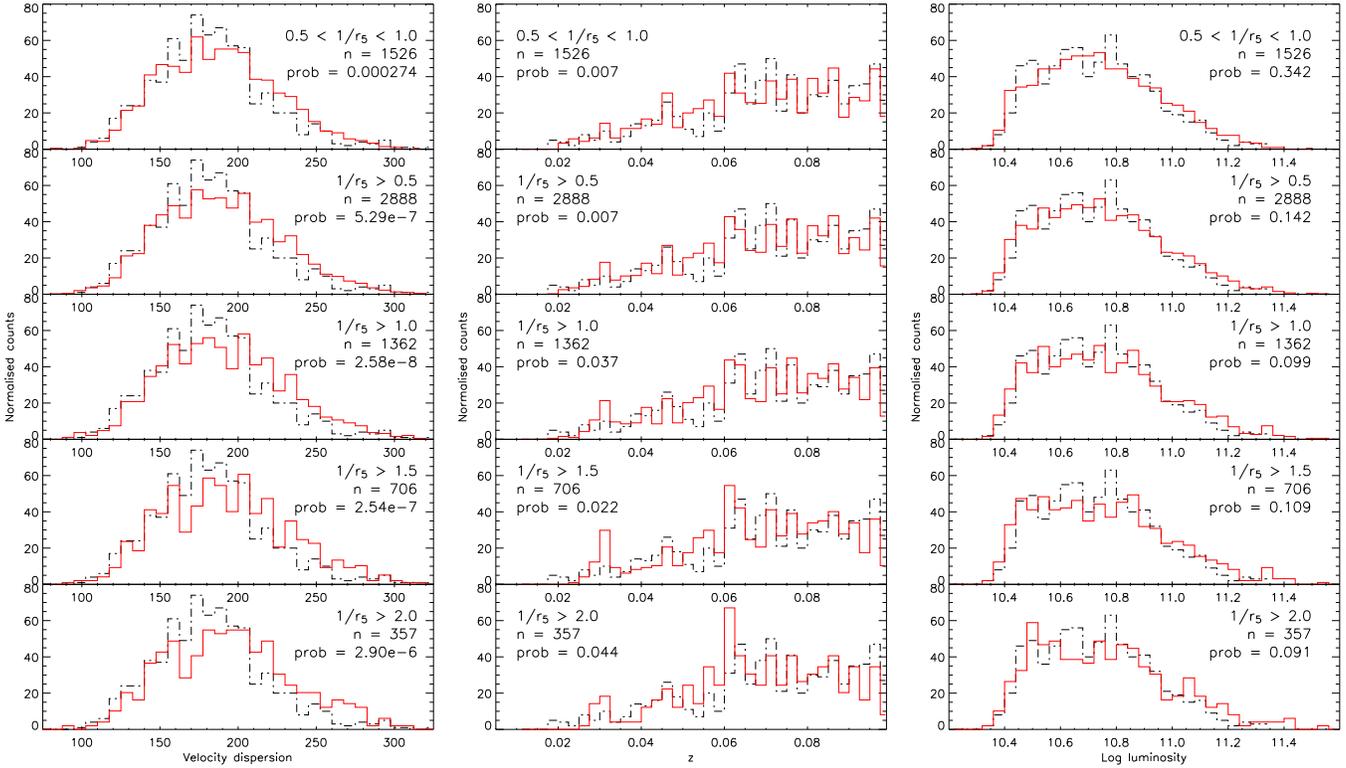}
\caption{Distribution of velocity dispersion, redshift and z-band luminosity 
for various density environments. In each case the dotted line is the 
distribution for the
comparison sub-sample with $1/r_5<0.5$ and the solid (red) line is that for
increasingly dense environments. Density increases from top to bottom in each
panel. The density of the environment considered, number of galaxies, and the KS-
probability that the 2 distributions are drawn from the same parent population
are shown in each plot. {\bf Left:} Velocity dispersion, {\bf middle:} redshift,
{\bf right:} luminosity. The clear increase of the velocity dispersion with
density of environment is quantified in Fig.~\ref{fig:PDens_sigma}}
\label{fig:histograms}
\end{figure*}

\begin{figure}
\includegraphics[clip,width=0.95\columnwidth]{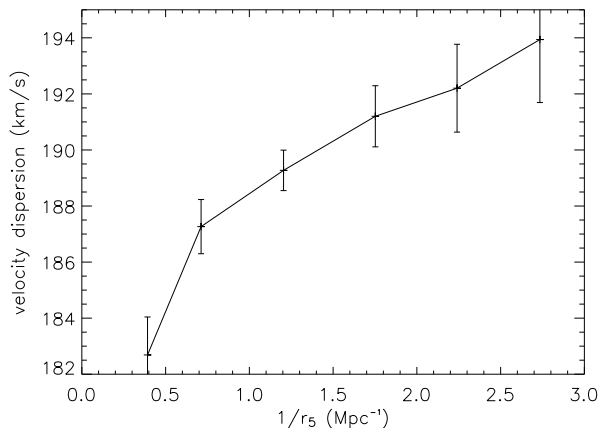}
\caption{Mean value of the velocity dispersion, $\sigma$, as a function of our
measure of the environmental density, $1/r_5$. The error bars are the
standard deviation of the distribution of velocity dispersion divided by the
square root of the number of objects.}
\label{fig:PDens_sigma}
\end{figure}

In Fig.~\ref{fig:histograms} we show the distributions of velocity dispersion,
redshift and luminosity for each of the sub-samples compared with the sub-sample
with $1/r_5 < 0.5$ along with the probability that the two distributions are
drawn from the same parent population. We note that the luminosity and redshift
distributions are similar in each case (although the two lowest density bins for 
the redshift distribution have a probability of only $0.7\%$ of being drawn from 
the same parent population their means are similar). 
The distributions of velocity dispersion on the other hand are very
significantly different in all cases. The velocity dispersion tends to be higher
in denser environments and increases monotonically with our measure of the
density of environment. At $1/r_5 < 0.5$,  $<$$\sigma$$> = 182.7\;\rm km\,s^{-1}$,
whereas for $1/r_5 > 2.0$, $<$$\sigma$$> = 193.9\;\rm km\,s^{-1}$. This trend is
illustrated in Fig.~\ref{fig:PDens_sigma}. In order to test that the differences in
velocity dispersion distributions are not due to a special distribution in the 
sub-sample with $1/r_5 < 0.5$ we also compared each environment with the sub-sample 
with $0.5<1/r_5<1.0$. Significant differences remained (although at lower significance
due to the smaller total number of objects and the smaller differences in $1/r_5$.

As we saw above, most indices show some correlation with
the velocity dispersion. The differences in mean velocity dispersion for
different environments therefore mean that differences in the index values
between environments will have a contribution due only to the index-$\sigma$
relation. The two effects might entirely explain any index differences we find
with environment. In order to take into account this effect we have employed a
Monte-Carlo approach.

Using the entire sample (before defining regions in different
environments) we consider the index-$\sigma$ relation. Binning the
data in $\sigma$ with a bin width of $10\;\rm km\,s^{-1}$ we plot the
value of the mean index per bin against $\sigma$ and linearly
interpolate between points.  This line is used to define the
index-$\sigma$ relation.

For a given sub-sample of environmental density we then take the
measured velocity dispersion value of each galaxy  and translate it
into a value for the index via the index-$\sigma$ relation. In the
process we alter the derived index value by a random amount drawn from
an appropriate Gaussian distribution so that the observed standard
deviation of measured indices is correctly reproduced. We thus create
a model data set with an index distribution defined only by the
velocity dispersion of the sub-sample and the index-$\sigma$ relation.

In this way we generate a model index distribution for a given
sub-sample of environmental density. Any difference between the index
values from two model sub-samples defined in this way is due only to
the difference in the distribution of velocity dispersion. We quantify
this difference by subtracting the means of two environmental density
sub-samples for 1000 realisations of the model. The distribution of
this model difference gives us a distribution against which to test
the observed difference in index. If the real data show a difference
well in excess of that seen in the simulated data then we identify an
effect of environment on the index.

\begin{figure*}
\includegraphics[clip,width=1.0\textwidth]{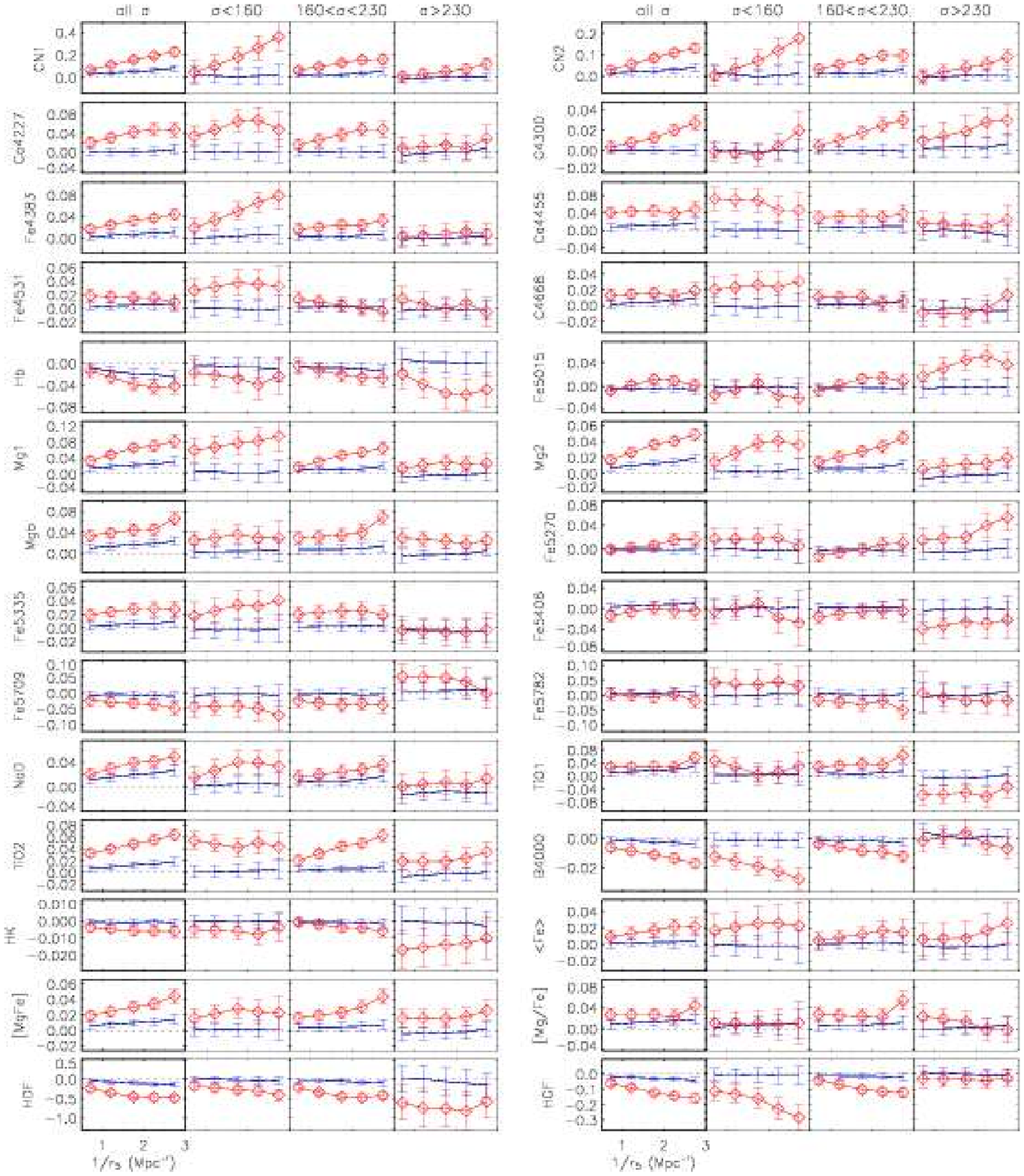}
\caption{Comparison of normalised index values with the comparison low-density sub-sample 
($1/r_5 < 0.5$) for increasingly dense environments. The horizontal axis shows
values of the mean density of the environment ($1/r_5$) and the vertical axis
shows the fractional deviation of the index value with respect to the sample
mean. Vertical error bars show the results of Monte-Carlo simulations which take into
account the index$-\sigma$ relation and the difference in the distribution of
velocity distribution between environments of different density. 
Open diamonds represent
the observed index difference. For each index four panels are shown. The {\bf left} 
panel shows results for the whole sample. The other 3 panels show the
results for 3 ranges of velocity dispersion. From left to right:
$\sigma < 160\;\rm km\,s^{-1}$, $160 \leq \sigma \leq 230\;\rm km\,s^{-1}$ and
$\sigma > 230\;\rm km\,s^{-1}$.   }
\label{fig:monte-together}
\end{figure*}

Fig.~\ref{fig:monte-together} displays the normalised index differences for 5
environments of increasing density, each compared to the sub-sample with the
lowest density ($1/r_5 < 0.5$). The normalised index difference is given by,

\begin{equation}
\Delta <I_s> = \frac{<I_s> - <I_{0}>}{<I>}
\end{equation}

\noindent where $<I_s>$ is the mean value of the index in the
sub-sample with mean environmental density $s$, $I_{0}$ is the mean
value of the index in the comparison sub-sample and $<I>$ is the mean
index value for the whole sample over all environmental
densities. Table~\ref{tab:monte-together} summarises the results of
Fig~\ref{fig:monte-together}.

\begin{table}
\caption{Comparison of normalised index values for the highest and lowest density 
sub-samples shown in the left panel of Fig.~\ref{fig:monte-together}. The data and 
the simulation refer to the whole sample. LDE and HDE refer to densities of $1/r_5 = 0.712$ 
and $1/r_5 = 2.734$.}
\label{tab:monte-together}
\begin{tabular}{@{}l@{\hspace{2mm}}c@{\hspace{2mm}}c@{\hspace{2mm}}c@{\hspace{2mm}}c}
\hline
Index & Simulation & Data  & Simulation & Data\\
         &     LDE        & LDE   &     HDE       &  HDE \\
\hline
       &  &  & & \\
CN1        &   0.034$\pm$0.020 & 0.066$\pm$0.028 & 0.083$\pm$0.026 &   0.227$\pm$0.036 \\
CN2        &   0.015$\pm$0.013 & 0.028$\pm$0.018 & 0.039$\pm$0.017 &   0.131$\pm$0.024 \\       
Ca4227  &   0.000$\pm$0.007 & 0.020$\pm$0.010 &  0.006$\pm$0.010 &   0.047$\pm$0.015 \\
G4300    &   0.000$\pm$0.003 & 0.003$\pm$0.005 & 0.000$\pm$0.004  &   0.027$\pm$0.007 \\
Fe4383   &   0.003$\pm$0.005 & 0.017$\pm$0.007 & 0.012$\pm$0.007 &   0.044$\pm$0.010 \\
Ca4455  &   0.008$\pm$0.008  & 0.041$\pm$0.011 & 0.015$\pm$0.012 &  0.048$\pm$0.016 \\
Fe4531   &   0.003$\pm$0.006  &  0.018$\pm$0.008 & 0.006$\pm$0.009 &  0.008$\pm$0.012 \\
C4668    &   0.002$\pm$0.005  &  0.013$\pm$0.007 & 0.009$\pm$0.007 & 0.019$\pm$0.010 \\
H$\beta$& -0.009$\pm$0.008 & -0.014$\pm$0.012 & -0.025$\pm$0.011 &  -0.041$\pm$0.015\\
Fe5015   & -0.003$\pm$0.005 & -0.008$\pm$0.007 & -0.007$\pm$0.007 &  0.002$\pm$0.010 \\
Mg1       &   0.014$\pm$0.007 &   0.032$\pm$0.010 & 0.031$\pm$0.0100  & 0.082$\pm$0.014\\
Mg2       &   0.007$\pm$0.004 &   0.017$\pm$0.006 & 0.019$\pm$0.005   &0.048$\pm$0.007 \\
Mgb       &   0.010$\pm$0.007 &  0.033$\pm$0.009  & 0.023$\pm$0.008 &   0.066$\pm$0.012\\
Fe5270  & -0.001$\pm$0.005 & -0.001$\pm$0.008 &-0.001$\pm$0.008 & 0.016$\pm$0.012\\
Fe5335  &   0.003$\pm$0.006 &   0.019$\pm$0.009 & 0.009$\pm$0.008 & 0.027$\pm$0.012 \\
Fe5406  &   0.004$\pm$0.008 & -0.014$\pm$0.011 & 0.010$\pm$0.012& -0.005$\pm$0.017\\
Fe5709  & -0.006$\pm$0.012 & -0.025$\pm$0.017 &-0.011$\pm$0.015 &-0.050$\pm$0.021 \\
Fe5782 &  0.007$\pm$0.014 & 0.003$\pm$0.020   & 0.014$\pm$0.017 & -0.020$\pm$0.025 \\
NaD       &  0.011$\pm$0.006 & 0.020$\pm$0.009   & 0.026$\pm$0.009 &  0.049$\pm$0.012 \\
TiO1      &  0.010$\pm$0.010 & 0.027$\pm$0.014   & 0.027$\pm$0.014 & 0.056$\pm$0.019 \\
TiO2      &  0.007$\pm$0.005 & 0.032$\pm$0.007   & 0.018$\pm$0.007 &0.064$\pm$0.009 \\
B4000   & -0.001$\pm$0.002 & -0.006$\pm$0.003 &-0.004$\pm$0.002 &-0.017$\pm$0.003 \\
HK        & -0.001$\pm$0.002 & -0.004$\pm$0.002 &-0.001$\pm$0.002 &-0.006$\pm$0.003 \\
<Fe>    &   0.001$\pm$0.006 &  0.009$\pm$0.008  & 0.004$\pm$0.008 &  0.022$\pm$0.012\\
$\rm [MgFe]$  &  0.006$\pm$0.004  &  0.019$\pm$0.006 & 0.014$\pm$0.006 &  0.044$\pm$0.008 \\
$\rm [Mg/Fe]$ &  0.009$\pm$0.008  &  0.027$\pm$0.011 & 0.018$\pm$0.010 &  0.044$\pm$0.015 \\
HdF      &-0.048$\pm$0.046  &-0.237$\pm$0.066 &-0.152$\pm$0.054 &-0.480$\pm$0.082 \\
HgF      &-0.019$\pm$0.015  &-0.062$\pm$0.022 &-0.042$\pm$0.019 &-0.157$\pm$0.027 \\
\hline
\end{tabular}
\end{table}

The following indices show a statistically significant density dependence in the
sense that index values are higher in more dense environments: CN1, CN2, Ca4227,
G4300, Fe4383, (Ca4455), Mg1, Mg2, Mgb, (Fe5335) NaD, TiO2, and [MgFe] (parentheses
indicate that we find an offset but not a trend with density). The B4000 index
shows the opposite trend with lower values in more rich environments but this is
simply due to how the index is defined in the SDSS. Although none of the
differences attributable to environment significantly increase the scatter of
index values of the population as a whole, the differences are very significant.

Indices which show an offset but no trend with density over the range of the plots of 
Fig.~\ref{fig:monte-together} show an excess of low values in the comparison low density
sub-sample.

The above indices have different sensitivities to age and metallicity
variations and, in general, an increase of a metal index may indicate
increased age and/or larger metallicity. As for the effects of the
metallicity, we recall that the sensitivity to separate elemental
variations is also important (enhancement) and has been investigated
by various authors (Tripicco \& Bell 1995, Trager et al, 1998, Thomas,
Maraston \& Bender, 2003,  Korn, Maraston \& Thomas 2005, Annibali et
al. 2005).  Indeed, from the indices in which an effect is seen with
density of environment (and from those which show little or no effect)
we can also attempt to identify elemental abundance variations. In
this respect carbon is of particular importance.

Tripicco \& Bell (1995) and Korn et al. (2005) showed that carbon has
a strong influence on many of the indices in which we find a
trend. However it is striking that  C4668, in spite of showing the
greatest sensitivity to carbon, displays at most only a marginal
increase with environmental density.  This is inconsistent with an
increase in carbon abundance being the cause of the differences we see
for these indices. It also suggests that, if a general abundance
increase is the cause of the index difference, carbon  is a bad tracer
of overall abundance and that its relative abundance should  decrease
as metallicity increases.

We note that we do not reproduce the results of  S\'anchez-Bl\'azquez
et al. (2003) who find both a  lower C4668 and CN2 index in for
early-type galaxies in the central  regions of the Coma cluster when
compared to the field. This study was based only on Coma galaxies and
the mean environmental density of galaxies in our sub-sample with
$1/r_{5} \geq 1.5$ is unlikely to be as high as for those at the
centre of the Coma cluster.

Besides the various dependencies of indices on elemental abundances
they are also, of course, affected by age. In order to make sense of
the trends of the indices we must make use of the sensitivity to the
different physical parameters simultaneously.

\subsection{Age, metallicity and $\alpha$-enhancement}
\label{sect:regression}
In order to extract trends in age, metallicity and $\alpha$-enhancement from our
measured trends in the line indices we have used the recent
models computed by Annibali (PhD thesis and Annibali et al. 2005)
and a multiple-linear regression analysis. 

These models account for the effects of age, metallicity and
$\alpha$-enhancement variations.  The models have been computed by
adopting the fitting functions that describe the dependence of the
Lick indices on stellar parameters. The effects of the element
partition has been introduced following the sensitivity to different
elements provided by  Korn et al. (2005). In computing the effects of
$\alpha$-enhancement the elements  N, O, Ne, Na, Mg, Si, S, Ca, Ti
were assigned to the {\it enhanced} group  while Cr, Mn, Fe, Co, Ni,
Cu, Zn to the {\it depressed} group.  Because our analysis makes use
of indices that show a mild to strong dependence on the relative
abundance of carbon, we have also considered the effects of enhancing
this element alone.

Our selection of indices to include in the regression analysis is
restricted to those satisfying two criteria. Firstly, the indices must
be well modelled by the SSPs of Annibali et al. (2005) and secondly,
those that have fractional errors $<20\,\%$. In common with similar
models by other workers (e.g. Thomas, Maraston \& Bender 2003),
certain indices are not reproduced well by the adopted SSP models. The
CN indices are underpredicted compared with globular cluster data
unless nitrogen is increased by a factor 3 with respect to the
$\alpha$-elements. NaD is sensitive to interstellar absorption
(Burstein et al. 1984) and so is of limited use for stellar population
studies (Worthey et al. 1994). The model predictions of the Ca4227
index are slightly too high  while Ca4455 suffers from calibration
problems (Maraston et al. 2003). The TiO1 and TiO2 indices appear to
be poorly calibrated, and furthermore measured TiO1 indices may be
affected by the presence of the broad NaD feature (Denicol\`o et
al. 2005). Among the Fe indices, we adopted the classic Fe5270 and
Fe5335 indices plus the blue Fe4383 and Fe4531 indces which are both
sensitive to iron and are well calibrated.The other iron indices are
relatively poorly calibrated (Fe5782), are less sensitive to Fe
(Fe5709), or may be affected by emission contamination (Fe5015,
Kuntschner et al. 2002). For the age-sensitive indicators we adopted
the classic H$\beta$ index, which is mostly independent of
$\alpha$-enhancement effects but may be affected by emission
contamination and we preferred H$\delta$ to H$\gamma$ as it is less
affected by nebular emission.

Our final list of indices is: H$\beta$, H$\delta$, Mg1, Mg2, Mgb,
Fe4383, Fe4531, Fe5270, Fe5335, G4300 and C4668. (We note that the
inclusion of H$\beta$ has no significant effect on the results. The
primary reason for it's inclusion was to allow comparison with the
results of other authors.) We find that our results are also not a
strong function of the adopted S/N cutoff of the spectra.

\begin{figure*}
{\centering \begin{tabular}{ccc}
\resizebox*{0.31\textwidth}{!}{\includegraphics{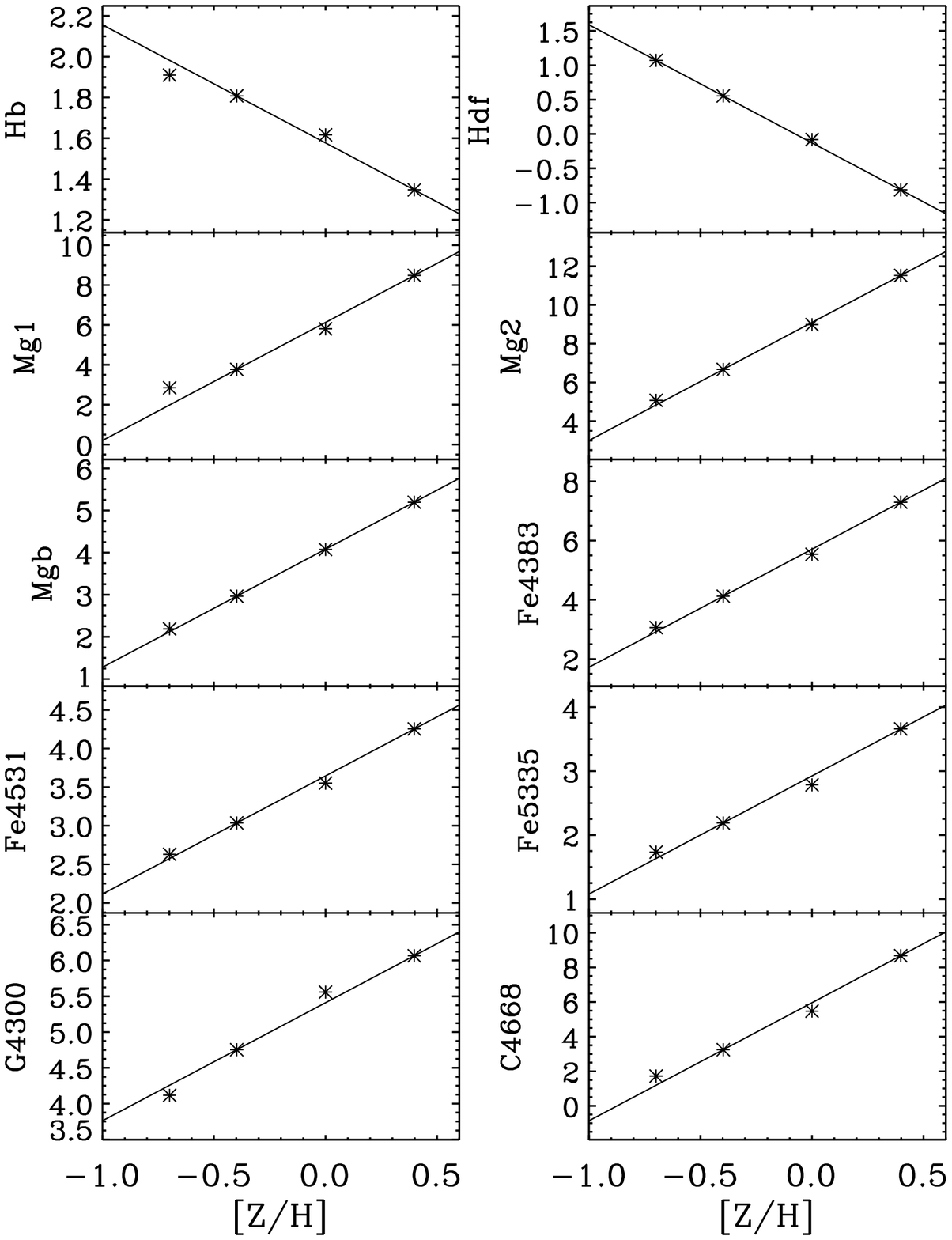}} &
\resizebox*{0.31\textwidth}{!}{\includegraphics{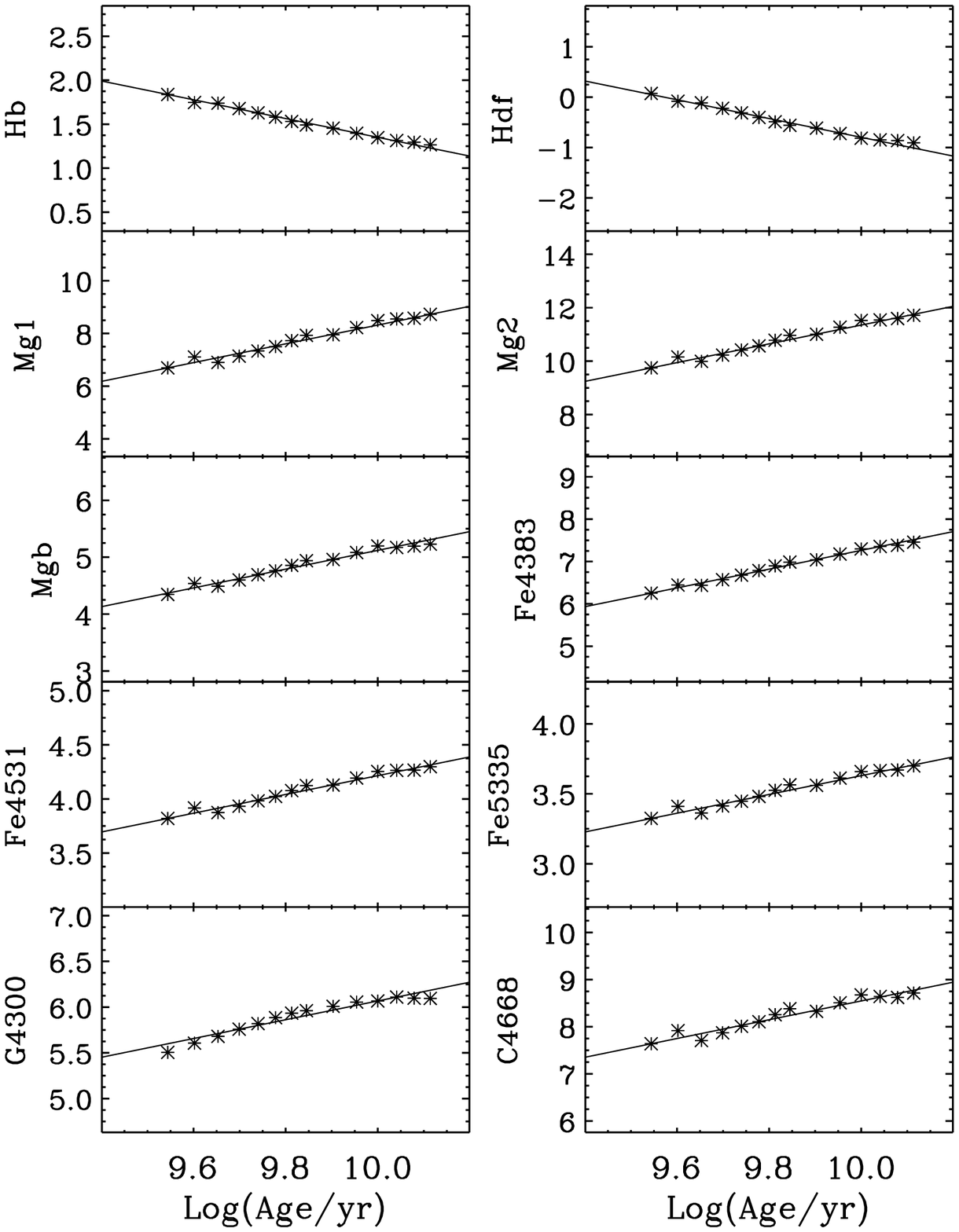}} &
\resizebox*{0.32\textwidth}{!}{\includegraphics{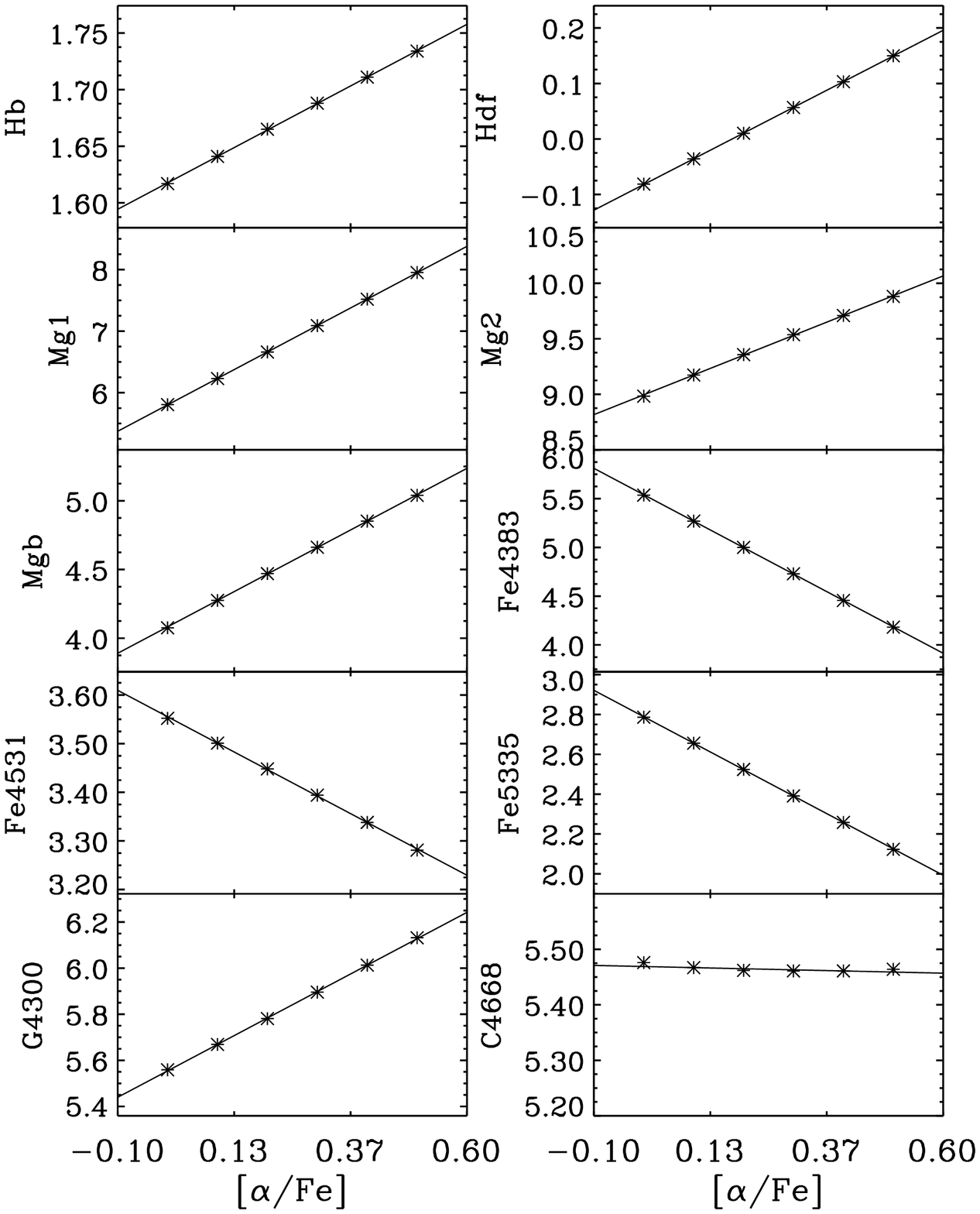}}
\end{tabular}}
\caption{The linearity of our adopted models for the selected indices. Index variations versus , {\bf left:} metallicity ($log(Z/Z_{\odot})$) at a fixed age of 10 Gyr and fixed $\rm [\alpha/Fe]$ and $\rm [C/H]$, {\bf middle:} $\log$(age), and {\bf right:} $\rm [\alpha/Fe]$.}   
\label{fig:linearity}
\end{figure*}

We express the observed index variation
as a linear combination of age, metallicity, $\alpha$-enhancement and carbon 
variation.
\begin{eqnarray}
\nonumber \lefteqn{\delta{I}=\frac{\partial I}{\partial \log(t)} \delta \log(t) + 
\frac{\partial I}{\partial \log(Z)} \delta \log(Z) + } \\
& & \frac{\partial I}{\partial [\alpha/{\rm Fe]}} \delta [\alpha/{\rm Fe}] + 
\frac{\partial I}{\partial [{\rm C/H}]} \delta [{\rm C/H}]
\label{eqreg}
\end{eqnarray}
where the partial derivatives are derived by the models and the
observed variation $\delta{I}$ is obtained with respect to a reference
point. Our use of a linear combination of model parameters is
justified in Fig.~\ref{fig:linearity} where we see that the index
variations are indeed linear with respect to changes in the model
parameters in equation~\ref{eqreg}.For a given observed data point, by
considering any number of indices larger than four at once,  we can
perform a multi-regression analysis and obtain the corresponding
variations of  $\delta$Log(t), $\delta$Log(Z), $\delta$[$\alpha$/Fe]
and $\delta$[C/H].  For the purposes of the regression analysis we
have first considered the whole sample regardless of environment and
then 2 sub-samples chosen to be representative of the `field' and
`cluster' environment. For the field sub-sample we take $1/r_{5} \leq 0.5$ 
and for the cluster sub-sample we take $1/r_{5} \geq 1.5$. We
then bin the index data in $\sigma$. Index variations are computed
relative to the index values for the whole sample at $\sigma = 200\;\rm km\,s^{-1}$.

\begin{figure*}

\includegraphics[angle=90,width=0.9\textwidth]{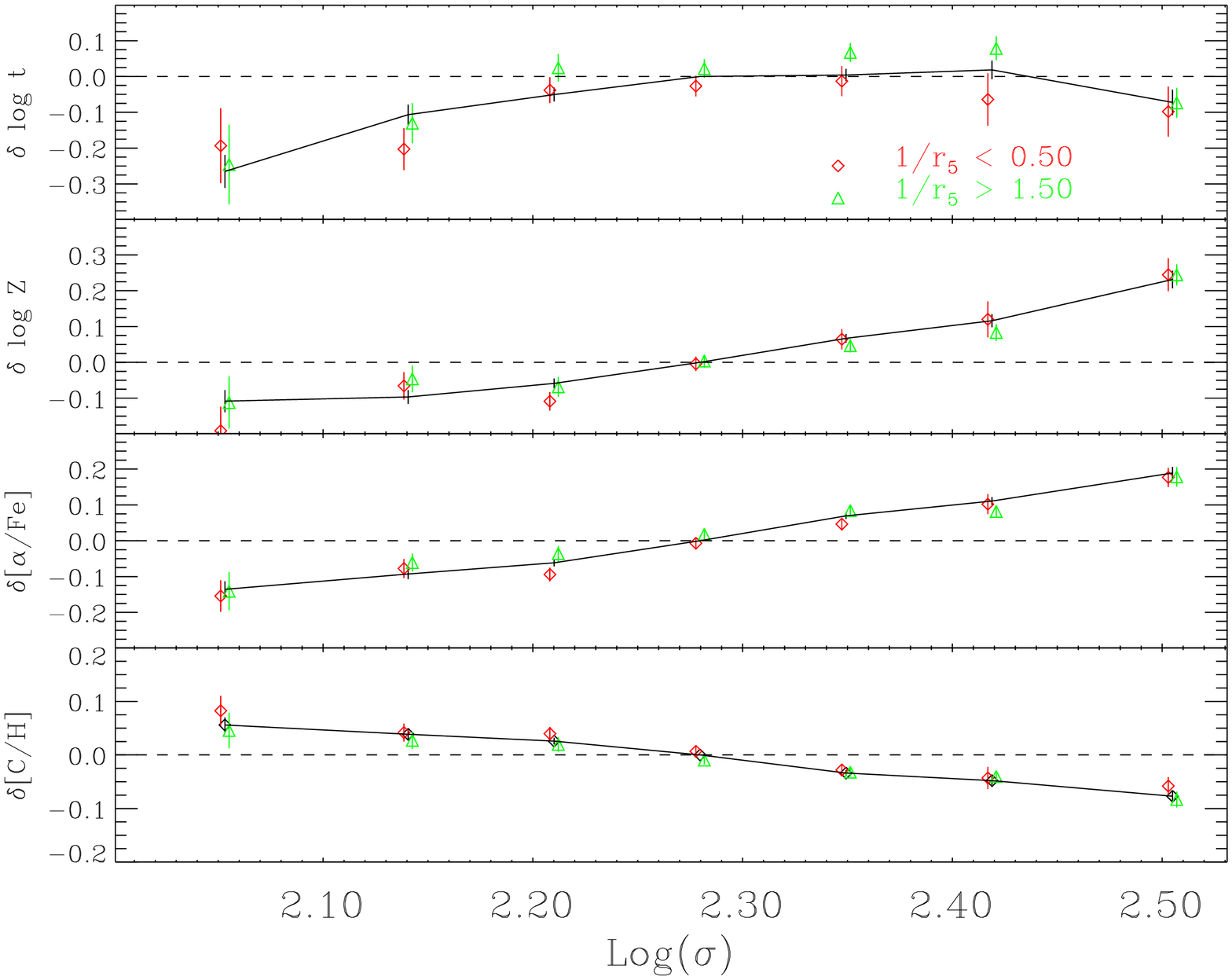}
\vskip 25pt 
\caption{Age, metallicity, $\alpha$-enhancement and carbon enhancement 
variations as
a function of velocity dispersion. A linear regression analysis has
been carried out simultaneously on the
H$\beta$, H$\delta$, Mg1, Mg2, Mgb, Fe4383, Fe4531, Fe5270, Fe5335, G4300 and C4668 indices.
The solid line represents
the entire sample, diamonds only those objects in low density environments
($1/r_5 < 0.5$) and triangles only those in high density environments
($1/r_5 > 1.5$). Values are differences
with respect to those of the entire
sample at a velocity dispersion of $200\;\rm km\,s^{-1}$. 
The number of objects in each bin in $\sigma$ is shown in Fig.~\ref{fig:regression_num}}
\label{fig:regression_new}
\end{figure*}

\begin{figure}
\includegraphics[angle=90,width=0.5\textwidth]{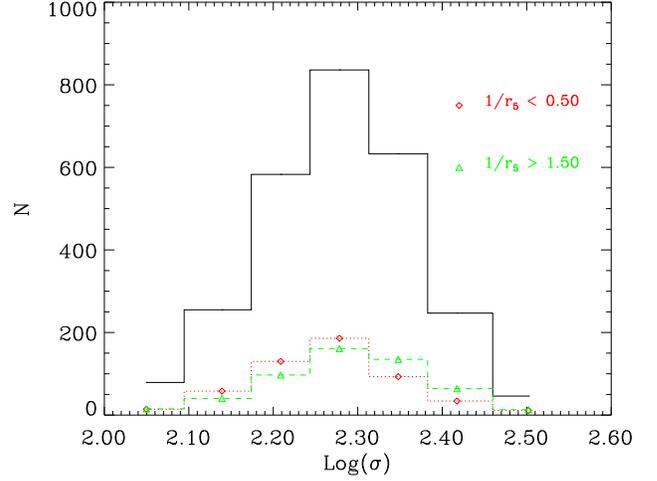}
\vskip 25pt 
\caption{The number of objects in each bin of Fig.~\ref{fig:regression_new}. 
The symbols have the same meanings as in Fig.~\ref{fig:regression_new}.}
\label{fig:regression_num}
\end{figure}

\begin{figure*}
\includegraphics[angle=90,width=0.95\textwidth]{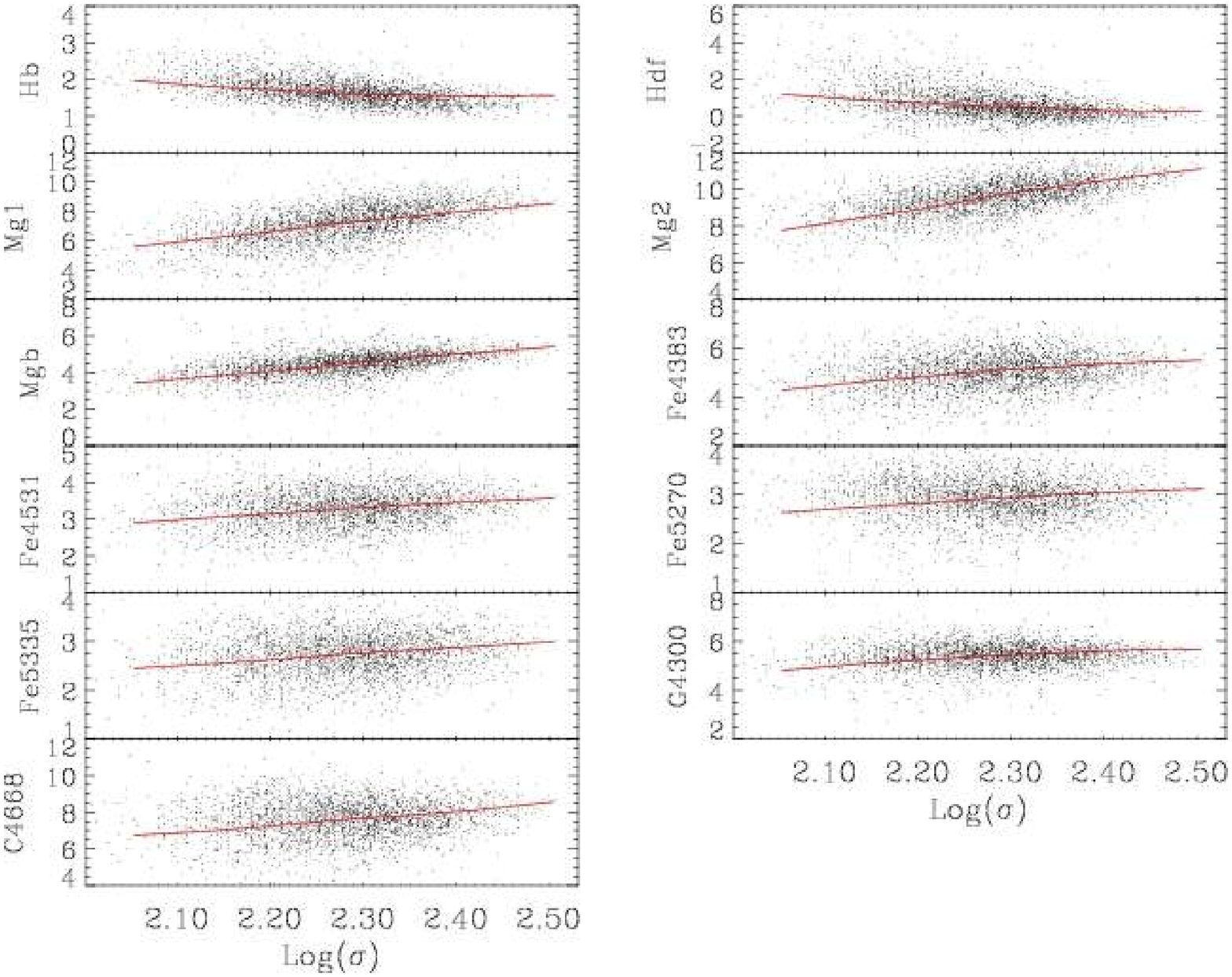}
\vskip 25pt 
\caption{A posteriori comparison of the models used in 
the regression analysis of Fig.~\ref{fig:regression_new} 
with the indices included in the analysis.
The lines are the simultaneous solution to the whole set of indices.}
\label{fig:regression_new_indices}
\end{figure*}

In a preliminary analysis we have not considered the variation
of the abundance of carbon  [C/H]. This allows a comparison of
the results obtained with our new models with those obtained
with the models of Thomas et al. (2004). The outcome of this
analysis can be summarised as follows. Although we always find an
$\alpha$-enhancement that increases with velocity dispersion our results for the
age and metallicity depend on which indices are included in the analysis. The
results are similar if we make use of the models of Thomas et al. (2004).
Specifically, our results change with the inclusion/exclusion of indices which
have a strong dependence on the carbon abundance such as C4668 and Mg1. As we
noted above the lack of strong C4668-$\sigma$ relation suggests that the index
does not trace overall metallicity any more than the mean iron abundance. The
different sensitivities specifically to carbon abundance of many of the indices
makes the consideration of this element of particular importance. 
We therefore subsequently
modified our analysis to include independently the effects of carbon enhancement.
Although the correction for carbon abundance is really just the most important
part of a correction that should take account of other elements such as titanium
and sodium we consider here the refinement only for carbon. As a consequence of
the modification, our results for the age and metallicity using the refined model
are much less dependent on the indices included in the regression analysis. With
this modified model we can include a larger number of indices and exclude only those
which have a relatively strong dependence on elements not explicitly modelled (
e.g. NaD, TiO1,2).

The results of this analysis are shown in
Figs.~\ref{fig:regression_new} and \ref{fig:regression_new_indices}.
Fig.~\ref{fig:regression_new} shows the run of the physical
parameters, expressed as differentials from the reference point.  In
this figure the solid line indicates the full sample while  triangles
and diamonds represent high and low density samples, respectively.
Fig.~\ref{fig:regression_new_indices} compares the values of the model
indices with the observed values, for the full sample.  The model
indices have been obtained by adding the model solutions found via
equation~\ref{eqreg} to the reference indices. In other words the
lines shown in  Figure~\ref{fig:regression_new_indices} represent the
simultaneous  solution to the whole set of indices.

We note immediately that only the age is influenced by the density of
environment; metallicity, $\rm [\alpha/Fe]$ and $\rm [C/H]$ show no
dependence on environment. We find the following trends with velocity
dispersion.

For the sample as a whole (solid line in
Fig.~\ref{fig:regression_new}) the age rises significantly from -0.25
dex at $100\;\rm km\,s^{-1}$ to a maximum near $260\;\rm km\,s^{-1}$. 
Neither Kuntschner et al. (2001) nor Annibali et
al. (2005) find any dependence of age on $\log(\sigma)$ whereas Thomas
et al. (2005) find a significant positive slope ($\simeq$0.24-0.32). 
Beyond $260\;\rm km\,s^{-1}$ there is then a slight fall
to -0.07 dex in the last bin at $320\;\rm km\,s^{-1}$, significant at
about the $2\sigma$ level. This decrease is more significant for
galaxies in dense environments. If this decrease is real it may be
evidence that the most massive galaxies in dense environments have
undergone some recent star formation.

In the range $140-260\;\rm km\,s^{-1}$ galaxies in low density
environments are $\simeq 0.07$ dex ($\sim 1.5\;\rm Gyr$ for an
elliptical of 10 Gyr) younger than their counterparts in dense
environments for a given $\sigma$. Although the significance of this
difference is low in any single bin in velocity dispersion the
consistency of the difference is evidence that it is a real
difference. This result is almost independent of the models adopted
and, since the comparison is performed at fixed $\sigma$, it is also
unlikely to be affected by the velocity dispersion correction or the
reference to the Lick system.

The metallicity shows a very significant, monotonic rise from -0.1 dex at 
$100\;\rm km\,s^{-1}$ to +0.25 at $320\;\rm km\,s^{-1}$ described well by,
\begin{equation}
\frac{d~\log(Z)}{d~\log(\sigma)}\simeq 0.76 
\label{zsig}
\end{equation}
\noindent and independent of environment. This is in good agreement
with the value found by Kuntschner et al. (2001). It is 20\% larger
than that derived by Annibali et al. (2005), 30\% larger than that found by Nelan et al. (2005) and 40\% larger than that of Thomas et al. (2005). Although in this latter case the
models show a zero-point offset in metallicity with respect to those
used in the present study this should not result in a different
gradient in the differential analysis carried out here. It is more
likely that the difference is due to differences in aperture or
velocity dispersion corrections. Bernardi et al. (2006) find values in the range 0.38 - 0.58, again somewhat lower than our value. 

The $\alpha$-enhancement rises from -0.15 dex to +0.2 dex from 100 to $320\;\rm km\,s^{-1}$. 
We fit the linear relation;
\begin{equation}
\frac{d~[\alpha/{\rm Fe}]}{d~\log(\sigma)}\simeq 0.74   
\label{alfasig}
\end{equation}
This slope is about 1.5 times that found by Annibali et al. (2005) and
about 2.5 times larger than that found by Thomas et al. (2005) and
Kuntschner et al. (2001). It is also a factor of 2 steeper than that found by Bernardi et al (2006) and Nelan et al. (2005). This chemical enrichment pattern shows that
the star formation process in low mass halos has occurred over a
longer time scale, showing that the effect of feedback must be
important in the baryon assembly process.  Whatever the details of
this feedback, environment has no role to play.

The carbon enhancement instead shows a steady fall from +0.05 dex to -0.075 
dex from the lowest to the highest velocity dispersion: 
\begin{equation}
\frac{d~[{\rm C/H}]}{d~\log(\sigma)}\simeq -0.31 
\label{csig}
\end{equation}
For more massive galaxies carbon is under-produced relative to other metals. This is the first time that such a trend can be quantified.

We see from Fig.~\ref{fig:regression_new_indices} that despite the trends seen with velocity dispersion in Fig.~\ref{fig:regression_new} there is nonetheless a large spread in the index values for any given value of $\sigma$. Therefore there exist galaxies with $\sigma \sim 100\;\rm km\,s^{-1}$ that are as old and metal rich as as those with $\sigma \sim 300\;\rm km\,s^{-1}$. However, the fact that the range of $\rm H\beta$ and $\rm H\delta$ values is much smaller at large velocity dispersion than at low velocity dispersion is evidence that the converse is not true: galaxies with $\sigma \sim 300\;\rm km\,s^{-1}$  are never as young and metal poor as the typical galaxy with $\sigma \sim 100\;\rm km\,s^{-1}$.

\section{Discussion}

The current sample of early-type galaxies has been selected in such a way as to
avoid the redshift/density bias that is introduced by traditional selection
techniques when applied to a survey with a limited sky coverage. With a sample so
defined we find that there is no difference in either the redshift distribution
or the z-band luminosity distribution of early-type galaxies in more or less
dense environments. We do however find that environment affects the distribution
of velocity dispersion, with the mean $\sigma$ rising with the local density of
the environment. This permits the analysis of the sample as a whole without
considering separate bins of either redshift or galaxy luminosity, and the
consideration only of the distribution of $\sigma$ within restricted ranges.
Furthermore the large number of galaxies allows a robust estimate of the 
mean values of the indices at increasing velocity dispersion, 
even when sub-samples of objects belonging
to different environments are selected. In fact the
presence of environmental effects has already been qualitatively inferred from 
the statistical analysis described in section ~\ref{enveffect}.

From the quantitative analysis performed with new models, 
in which we have included all the major dependencies from the physical parameters, 
we infer that both the enhancement of $\alpha$-elements and the global metallicity 
increase with galactic velocity dispersion.
Instead, contrary to previous studies, we find
that the age initially increases but then flattens, and possibly then falls at the 
highest velocity dispersions.

While some of these trends have already been uncovered by
previous studies (Bressan et al. 1996, 
Thomas et al., 2005) it is only by means of
a careful statistical analysis of a large data sample
and its interpretation with the most up-to-date SSPs presented here,
that it is possible to obtain perhaps the cleanest scenario so far
for the formation and evolution of early type galaxies in different environments.
Despite broad agreement with recent studies based on similarly large samples (Nelan et al., 2005, Bernardi et al., 2006, Smith et al., 2006) one key difference emerges; that is our finding that the \emph{only} environment sensitive parameter is age. The above authors also find slightly enhanced $\alpha$ abundances in denser environments. Whether this difference is a result of sample selection criteria or diverse approaches to data analysis is not clear.

The general picture which emerges from our work is that at fixed mass ($\sigma$), galaxies 
formed first in dense environments (that now correspond to clusters) and $\sim 1.5\;\rm Gyr$ later in the less dense field, but with a similar star formation time-scale. If the early collapse in high density peaks happened
at a redshift of say $z \geq 6$ then, a $1.5\;\rm Gyr$ delay is compatible with a
late collapse at redshift of $z \simeq 2.5$. Lower mass galaxies ($\sigma \simeq 100\;\rm km\,s^{-1}$) formed $\sim 4\;\rm Gyr$ more recently in both environments, which would correspond to a mean formation redshift of $\sim 1.1$ if those with $\sigma \simeq 220\;\rm km\,s^{-1}$ formed at $z=6$. The star formation rate that characterised the mass assembly was more rapid for more massive objects (from the increase in $\rm [\alpha/\rm Fe]$ with $\sigma$).

The observations are at odds with the predictions of usual
hierarchical models of galaxy formation, where smaller halos typically
collapse earlier than the larger ones (Lacey \& Cole, 1994),
indicating that an important ingredient is still missing in these
models.  If we assume a standard initial mass function (IMF) i.e. not
biased toward high mass stars,  the trend of $\alpha$-enhancement with
velocity dispersion suggests that the stars that make up more massive
early-type galaxies formed over characteristically shorter timescales
than those that make up lower mass objects.  Furthermore, in the
absence of an effect of environment on the $\alpha$-enhancement,  we
are led to conclude that the timescale of a typical star formation
episode is not affected directly by the environment, only by the mass
of the object in question.  Since at constant efficiency the metal
enrichment is larger when  the star formation process lasts longer,
this implies that  {\sl the baryon assembly must have been much less
efficient  in smaller systems}.  This is strong evidence for the
importance of the energetic feedback on the star formation process, as
recently shown by  Granato et al. (2004). In these models the star
formation process in smaller systems  is rendered less efficient by
the supernova feedback with the result that, though starting earlier,
the star formation  lasts longer, giving rise to a lower metal
enrichment.

Although slower star formation with stronger feedback in low-mass
systems explains the observed trends in metallicity and
$\alpha$-enhancement it does not get over the problem of how most of
the star formation occurred after the mass was assembled. If this were
not the case we would not observe a rising $\rm [\alpha/Fe]$ with
velocity dispersion. Given that gas-rich mergers are expected to be
accompanied by bursts of star formation it is difficult to see how the
star formation can be delayed if multiple-mergers are responsible for
the mass assembly. Conversely, a small number of mergers would produce
systems with a significant net angular momentum.

It is not possible from the present analysis to quantify the  effects
of major mergers, though their efficiency should be higher in the kind
of low-velocity encounter typical of the field than in the
high-velocity encounters expected in clusters. It could then also be
possible that the formation process in field early type galaxies  is
modulated by the occurrence of major mergers. Annibali et al. (2005)
have shown that in that case the fraction of stars formed during the
burst is less than 25\%.

Finally, we find that [C/H] decreases at increasing metal content. 
This carbon anomaly is suggested by comparison of observed index values
(Fig.~\ref{fig:monte-together}) and is thus independent of the models.

Models in which [C/H] scales with the metal content do not
provide consistent results between indices with different sensitivity
to this element. The variation of [C/H] between low and high $\sigma$
galaxies is approximately 0.125 dex, independent of environment.
It is worth noting that several indices, usually adopted
for this kind of theoretical analysis, have a more or less pronounced
dependence on the carbon abundance (including
Mg1, Mg2 and to a lesser extent Mgb, see e.g. Korn et al. 2005).
It is thus important to perform the analysis with
suitable models that account for this effect.
This abundance trend may suggest a lower production of carbon in more metal rich, high
mass stars. But it may also indicate a lower C yield from intermediate
mass stars as the `hot bottom burning' process becomes more efficient
at increasing metal content (Marigo, Bressan \& Chiosi 1998)

\section{Conclusions}

We have carried out a detailed analysis of the spectral line indices
of a sample of 3614 early-type galaxies selected form the Sloan
Digital Sky Survey. Our selection technique avoids the
redshift/environment bias introduced by traditional selection criteria
when applied to a survey with partial sky coverage. We have quantified
the density of environment for each object and have investigated the
effects of environment on age, metallicity and $\alpha$-enhancement
via a regression analysis of the whole sample. Because the carbon
abundance has a strong effect on the strength of many indices we have
used revised models which take account also of the carbon abundance
explicitly. Thus, our models  take into account the main parameters
that drive the set of indices adopted in the analysis; age,
metallicity, $\alpha$-enhancement and carbon content.

Our results for early-type galaxies can be summarised as follows.

\begin{itemize}
\item Neither the z-band luminosity nor the redshift distribution are a 
function of galaxy environment out to a redshift of 0.1.

\item The mean velocity dispersion is higher in denser environments.

\item The only effect of environment is on the mean age of early-type galaxies 
for a given velocity dispersion. Environment has no effect on metallicity, 
$\alpha$-enhancement or [C/H].

\item The star formation episodes responsible for the assembly of the baryonic mass of
early-type galaxies were of shorter duration in objects which have a higher
present day mass. Environment did not influence the duration of these episodes.

\item The youngest objects are those with the lowest masses, both in
high and low density environments.

\item Objects in the field with all but the highest velocity dispersion are, on average, $1.5\;\rm Gyr$ younger
than their cluster counterparts. The same trend is also obtained by adopting the models by Thomas et al. (2005).  

\item Carbon becomes increasingly under-abundant in higher mass objects.

\end{itemize}

The trend we find of increasing $\alpha$-enhancement with galaxy mass
is consistent with a pure monolithic collapse model. However, the more
prolonged star formation in lower mass objects should also produce a
higher metallicity in these objects, which is not seen. A pure
monolithic collapse model is therefore rejected. In a simple
hierarchical merger scenario the $\alpha$-enhancement trend is not
reproduced. The $\alpha$-enhancement trend could be explained in a
hierarchical model if small mass halos which formed within a large
dark matter halo formed their stars more rapidly than equivalent mass
objects in a smaller parent halo. However, such a picture would
predict a dependency of the star formation timescale (and thus
$\alpha$-enhancement) also on environment. We see no such effect with
environment. Galaxy formation in which the star formation process is
dependent only on the mass of the galaxy is not consistent with either
simple monolithic collapse or hierarchical merging.

A rising $\alpha$-enhancement and metallicity with velocity dispersion 
requires a feedback mechanism that results in, 1) a slower star formation 
in low mass objects, and 2) a less efficient star formation in low mass objects.
The latter ensures that the more prolonged star formation in low mass
objects does not result in more metals. A model with mass-dependent
feedback has been investigated by Granato et al. (2004)

Small mass halos then, can collapse before larger mass halos, only if
the star formation within them is inhibited by such a  feedback
mechanism, such that, in the absence of mergers, their star formation
proceeds at an extremely slow rate. When mergers take place the
increased mass reduces the feedback and the rate and efficiency of
star formation increases. The greater the merger mass the more the
star formation rate is increased and the sooner it ends. The result is
that the mean age of the stars formed increases with mass even though
the merger epoch can be earlier for lower mass systems. There is
therfore an important distinction between the epoch of assembly of the
dark halo and the mean age of the stars in a galaxy which we measure
spectroscopically. The difference is largest for low mass galaxies. In
environments with a high density the merger process occurred $\sim
1.5\;\rm Gyr$ earlier than in low density environments and this timing
seems to be the only influence of the environment.

\section{Acknowledgements}
M.C. thanks the Istituto Nazionale di Astrofisica for support in the
form of a Research Fellowship. We also thank L. Danese for useful
discussions. We are grateful to an anonymous referee who's prompt comments
led to a substantial improvement in the article.

Funding for the creation and distribution of the SDSS Archive has been
provided by the Alfred P. Sloan Foundation, the Participating
Institutions, the National Aeronautics and Space Administration, the
National Science Foundation, the U.S. Department of Energy, the
Japanese Monbukagakusho, and the Max Planck Society. The SDSS Web site
is http://www.sdss.org/.

The SDSS is managed by the Astrophysical Research Consortium (ARC) for
the Participating Institutions. The Participating Institutions are The
University of Chicago, Fermilab, the Institute for Advanced Study, the
Japan Participation Group, The Johns Hopkins University, the Korean
Scientist Group, Los Alamos National Laboratory, the Max-Planck-
Institute for Astronomy (MPIA), the Max-Planck-Institute for
Astrophysics (MPA), New Mexico State University, University of
Pittsburgh, University of Portsmouth, Princeton University, the United
States Naval Observatory, and the University of Washington.

\section{References}
\bibliographystyle{mn2e} 
\bibliography{../../galevol.bib}
Annibali F., Rampazzo R., Bressan A., Danese L., Bertone E., Chavez M., Zeilinger W.W., 2005, astro-ph/0501302\\
Balogh M., Eke V., Miller C., et al., 2004, MNRAS, 348, 1355\\
Balogh M.L., Morris S.L., Yee H.K.C., Carlberg R.G., Ellingson E., 1999, ApJ, 527, 54\\
Bender R., Burstein D., Faber S.M., 1993, ApJ, 411, 153 \\
Bernardi M., Sheth R.K., Nichol R.C., Schneider D.P., Brinkmann J., 2005a, AJ, 129, 61 \\
Bernardi M., Nichol R.C., Sheth R.K., Miller C.J., Brinkmann J., 2006, AJ, 131, 1288\\
Bernardi M., Renzini A., da Costa L.N., Wegner G., Alonso M.V., Pellegrini P.S., Rit\'e C., Willmer C.N.A., 1998, ApJ, 508, L143\\
Bernardi M. et al. 2003, AJ, 125, 1882\\
Bower R.G., Lucey J.R.,  Ellis R.S., 1992, MNRAS, 254, 601 \\
Bressan A., Chiosi C., Tantalo R., 1996, A\&A, 311, 425\\
Burstein D., Faber S.M., Gaskell C.M., Krumm N., 1984, ApJ, 287, 586\\ 
Cappellari M., Bacon R., Bureau M., Damen M.C., Davies R.L., de Zeeuw P.T., Emsellem E., Falcon-Barroso J., Krajnovic D., Kuntschner H., McDermid R.M., Peletier R.F., Sarzi M., van den Bosch R.C.E., van de Ven G., 2006, MNRAS, 366, 1126\\
Croton D.J., Farrar G.R., Norberg P. et al., 2005, MNRAS, 356, 1155\\
Denicol\'o, G., Terlevich R., Terlevich E., Forbes D.A., Terlevich A., 2005, MNRAS, 358, 813\\
Dressler A., 1980, ApJ, 236, 351\\
Fukugita M., \& Peebles P.J.E., 2004, ApJ, 616, 643\\
G\'omez P.L., Nichol R.C., Miller C.J., Balogh M.L., Goto T., Zabludoff I., Romer A.K., Bernardi M., Sheth R., Hopkins A.M., Castander F.J., Connolly A.J., Schneider D.P., Brinkmann J., Lamb D.Q., SubbaRao M., York D.G., 2003, ApJ, 584, 210\\
Gonz\'alez J., 1993, Ph.D. thesis, University of California, Santa Cruz\\
Goto T., Yamauchi C., Fujita Y., Okamura S., Sekiguchi M., Smail I., Bernardi M., G\'omez P., 2003, MNRAS, 346, 601\\
Granato G. L., De Zotti G., Silva L., Bressan A., Danese L., 2004, ApJ, 600, 580\\ 
Gunn J.E., \& Gott R., 1972, ApJ, 176, 1\\ 
Hamilton D., 1985, ApJ, 297, 371\\
Hogg D.W., Blanton M.R., Brinchmann J., Eisenstein D.J., Schlegel D.J., Gunn J. E., McKay T.A., Rix H-W., Bahcall N.A., Brinkmann J., Meiksin A., 2004, ApJ, 601, L29\\
Hogg D.W., Blanton M.R., Eisenstein D.J., Gunn J.E., Schlegel D.J., Zehavi I., Bahcall N.A., Brinkmann J., Csabai I., Schneider D.P., Weinberg D.H., York D.G., 2003, ApJ, 585, L5\\
Jorgensen I., 1997, MNRAS, 288, 161\\
Korn A.J., Maraston C., Thomas D., 2005, A\&A, 438, 685\\
Kuntschner H., Smith R.J., Colless M., Davies R.L., Kaldare R., Vazdekis A., 2002, MNRAS, 337, 172\\
Kuntschner H., Lucey J.R., Smith R.J., Hudson M.J., Davies R.L., 2001, MNRAS, 323, 615\\
Lacey C., Cole S., 1994, MNRAS, 271, 676\\
Longhetti M., Bressan A., Chiosi C., Rampazzo R., 2000, A\&A, 353, 917\\ 
Maraston C., Greggio L., Renzini A., Ortolani S., Saglia R.P., Puzia T.H., Kissler-Patig M., 2003, A\&A, 400, 823\\
Marigo P., Bressan A., \& Chiosi C., 1998, A\&A, 331, 564\\ 
McIntosh D.H., Rix H-W., Caldwell N., 2004, ApJ, 610, 161\\
Miller C.J., Nichol R.C., G\'omez P.L., Hopkins A.M., Bernardi M., 2003, ApJ, 597, 142\\
Moore B., Katz N., Lake G., Dressler A., Oemler Jr. A., 1996, Natur, 379, 613\\
Nelan J. E., Smith R. J., Hudson M. J., Wegner G. A., Lucey J. R., Moore S. A. W., Quinney S. J., Suntzeff N. B., 2005, ApJ, 632, 137\\
Nikolic B., Cullen H., Alexander P., 2004, MNRAS, 355, 874\\
Puzia T.H., Saglia R.P., Kissler-Patig M., Maraston C., Greggio L., Renzini A., Ortolani S., 2002, A\&A 395, 45\\
Rampazzo R., Annibali F., Bressan A., Longhetti M., Padoan F., Zeilinger W.W. 2005, A\&A 433, 497\\
Rose J.A., 1984, AJ, 89, 1238\\
Rose J.A., 1985, AJ, 90, 1927\\
S\'anchez-Bl\'azquez, P., Gorgas J., Cardiel N., Cenarro J., Gonz\'alez J.J., 2003, ApJ, 590, L91\\
Shimasaku K., et al., 2001, AJ, 122, 1238\\
Smith R.J., Hudson M.J., Lucey J.R., Nelan J.E., Wegner G.A., 2006, astro-ph/0603688\\
Strauss M.A., et al., 2002, AJ, 124, 1810\\
Thomas D., Maraston C., Bender R., de Oliveira C.M., 2005, ApJ, 621, 673\\
Thomas D., Maraston C., Korn A., 2004, MNRAS, 351, L19\\
Thomas D., Maraston C., Bender R., 2003, MNRAS, 339, 897\\
Trager S.C., Faber S.M. Worthey G., Gonz\'alez J.J., 2000, AJ, 120, 165\\
Trager S.C., Worthey G., Faber S.M., Burstein D., Gozalez J.J., 1998, ApJS, 116, 1\\
Tripicco M.J., Bell R.A., 1995, AJ, 110, 3035\\
Worthey G., \& Ottaviani D.L., 1997, ApJS, 111, 377\\
Worthey G., Faber S.M., Gonzalez J.J., Burstein D., 1994, ApJS, 94, 687\\

\appendix

\section{Robustness of results to velocity dispersion correction of indices}
\begin{figure*}
\includegraphics[clip,width=16cm]{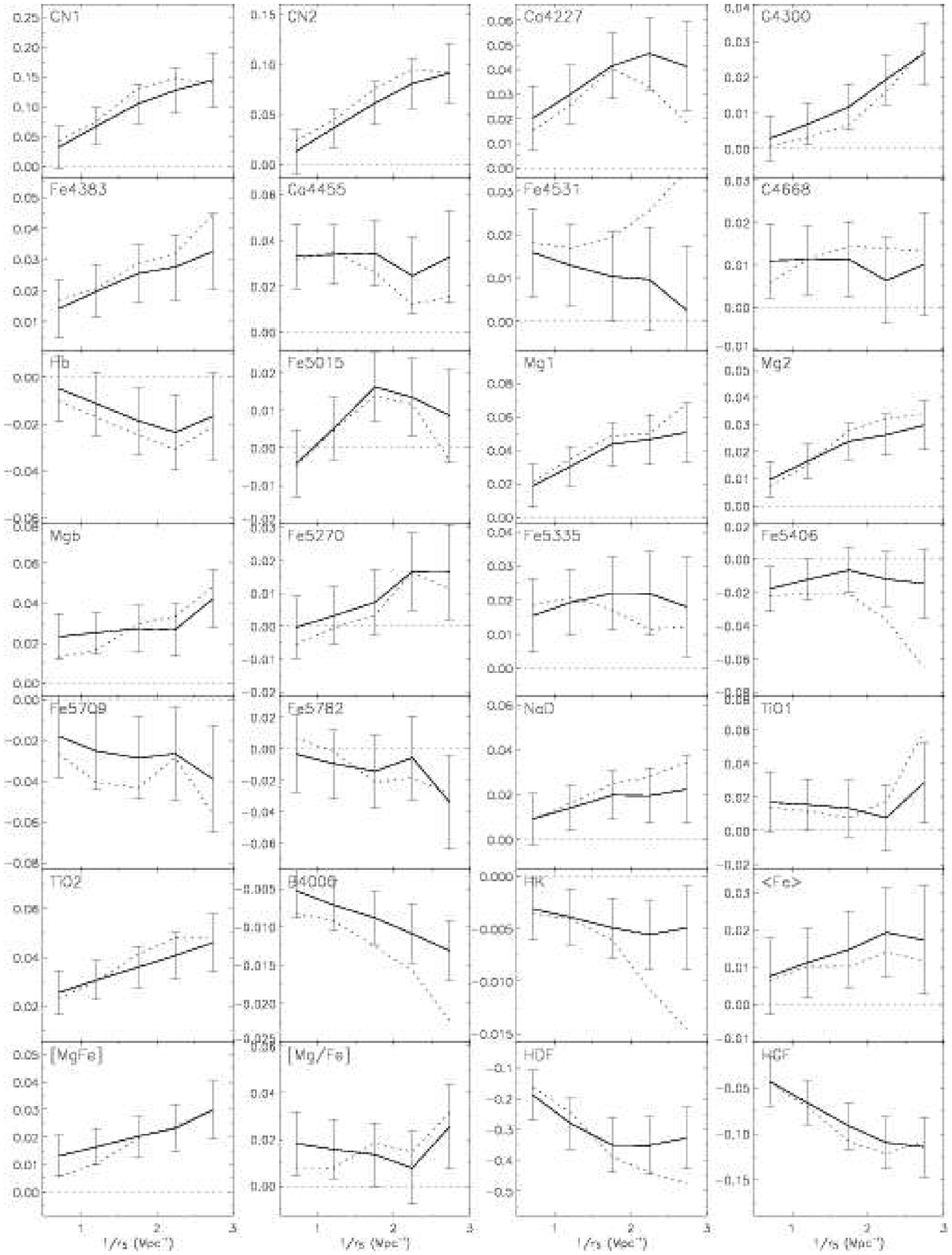}
\caption{Effect of environment on relative line index strength free of the effect
of the index-$\sigma$ relation. Similar to Fig~\ref{fig:monte-together} but shown
is the difference between the data and simulation. The solid line is for the the
sub-samples exactly as defined in Fig~\ref{fig:monte-together} and the dotted
line is for sub-samples which have been constrained (by random sampling in bins
of $\sigma$) to have similar velocity dispersion distributions. The error bars
are not shown for the second line but are similar to those of the solid line. The
similarity of the lines shows that neither an error in velocity dispersion
correction nor the differences in velocity dispersion distribution between sub-
samples can explain the effect we find with environment.}
\label{fig:simtest}
\end{figure*}

Our findings, that the index values for early-type galaxies are affected by
environment, are not due to an error in our velocity dispersion correction. If an
error in velocity dispersion correction introduced an erroneous gradient
component into the index-$\sigma$ relations
this would not alter the separation between the 2 lines in Fig.~\ref{fig:monte-together} 
because the Monte-Carlo simulations use the \emph{measured} index-$\sigma$ relations 
in order to establish how values of $\sigma$ are translated
into values of the index. In order to show this we carried out two tests. In the
first we introduced an error term into the gradient of the index-$\sigma$
relations and re-computed the simulations illustrated in fig.~\ref{fig:monte-together}. 
No differences were seen that extended beyond the error bars in this
figure. Secondly, we constrained the $\sigma$-distributions of the sub-samples in
each environment to be identical to that of the comparison sub-sample (
$1/r_5 <0.5$) by randomly selecting objects until the number of objects per bin
was in constant proportion to the numbers in the comparison sample. This of
course leads to the exclusion of many objects but as this is at random only the
size of the errors on individual points will be affected. Reproducing 
Fig.~\ref{fig:monte-together} in this way also did not change the index difference between
data and simulations. The results of this test are shown in
Fig.~\ref{fig:simtest} where we have subtracted the simulation from the data to
illustrate the effect of environment alone on the line indices. In short, our
results on the effect of environment are robust to errors in velocity dispersion
correction because it is not dependent on the form of the index-$\sigma$
relation.

\begin{figure*}
{\centering \begin{tabular}{ccc}
\resizebox*{0.33\textwidth}{!}{\includegraphics{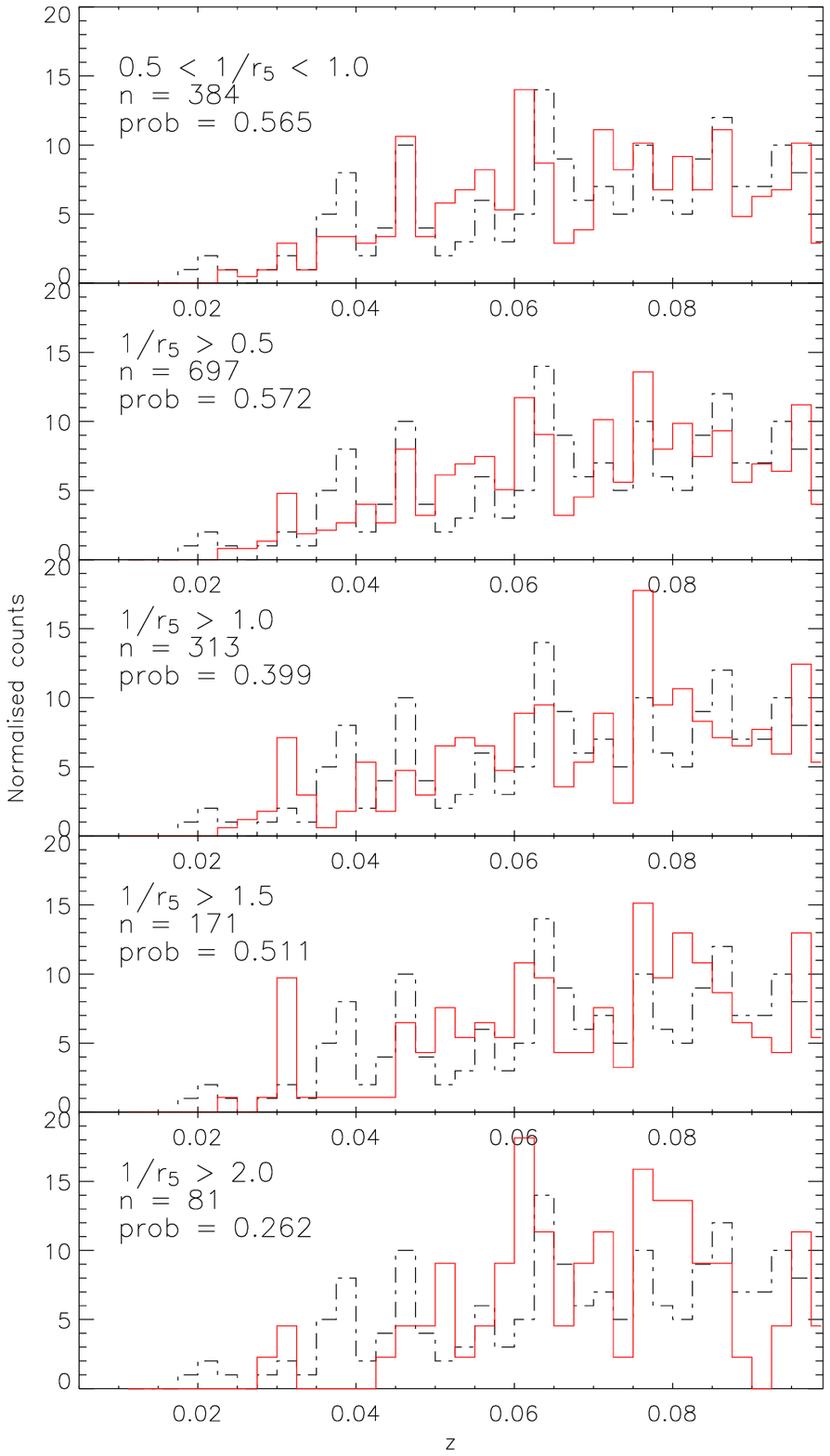}} &
\resizebox*{0.33\textwidth}{!}{\includegraphics{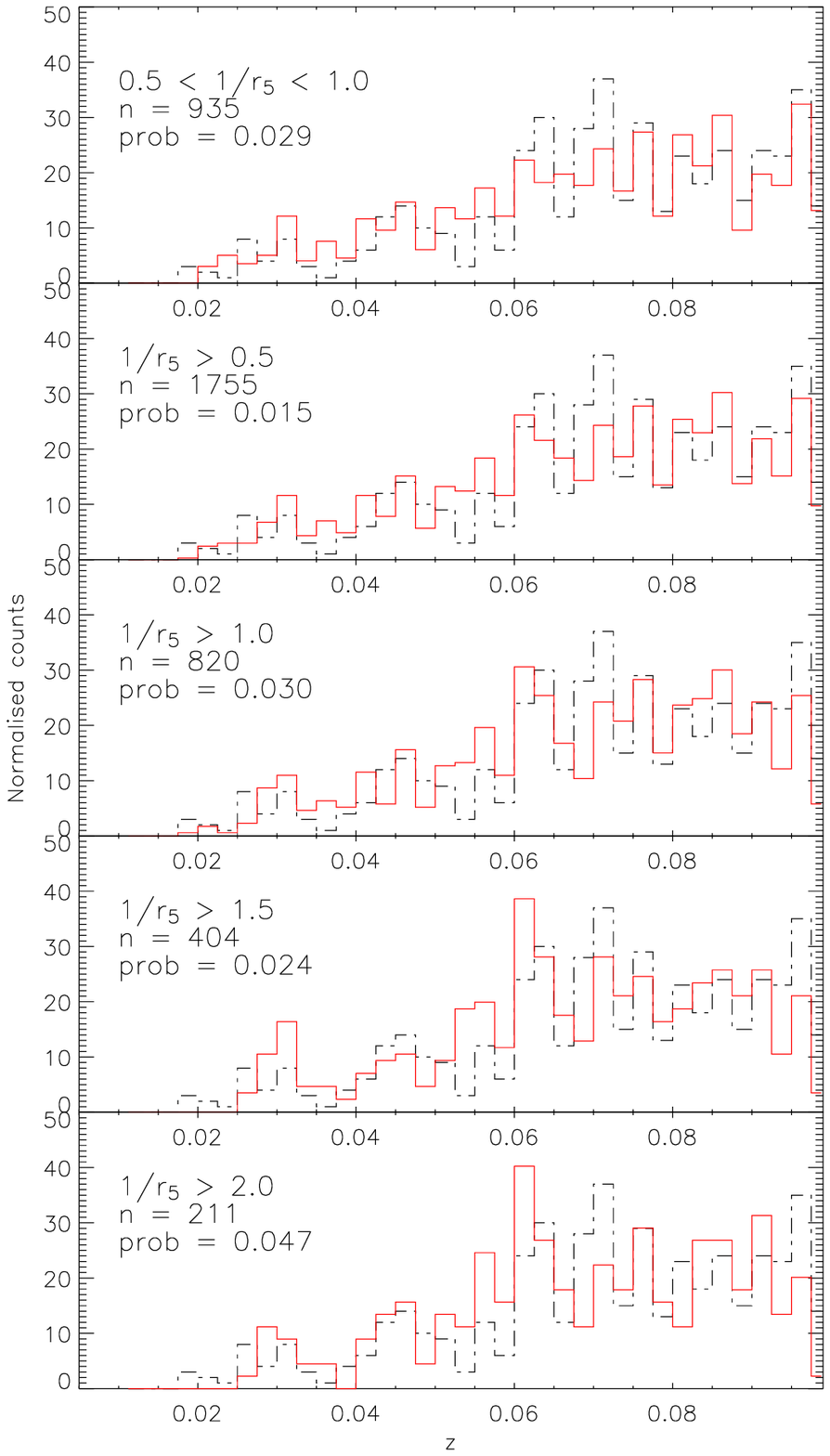}} &
\resizebox*{0.33\textwidth}{!}{\includegraphics{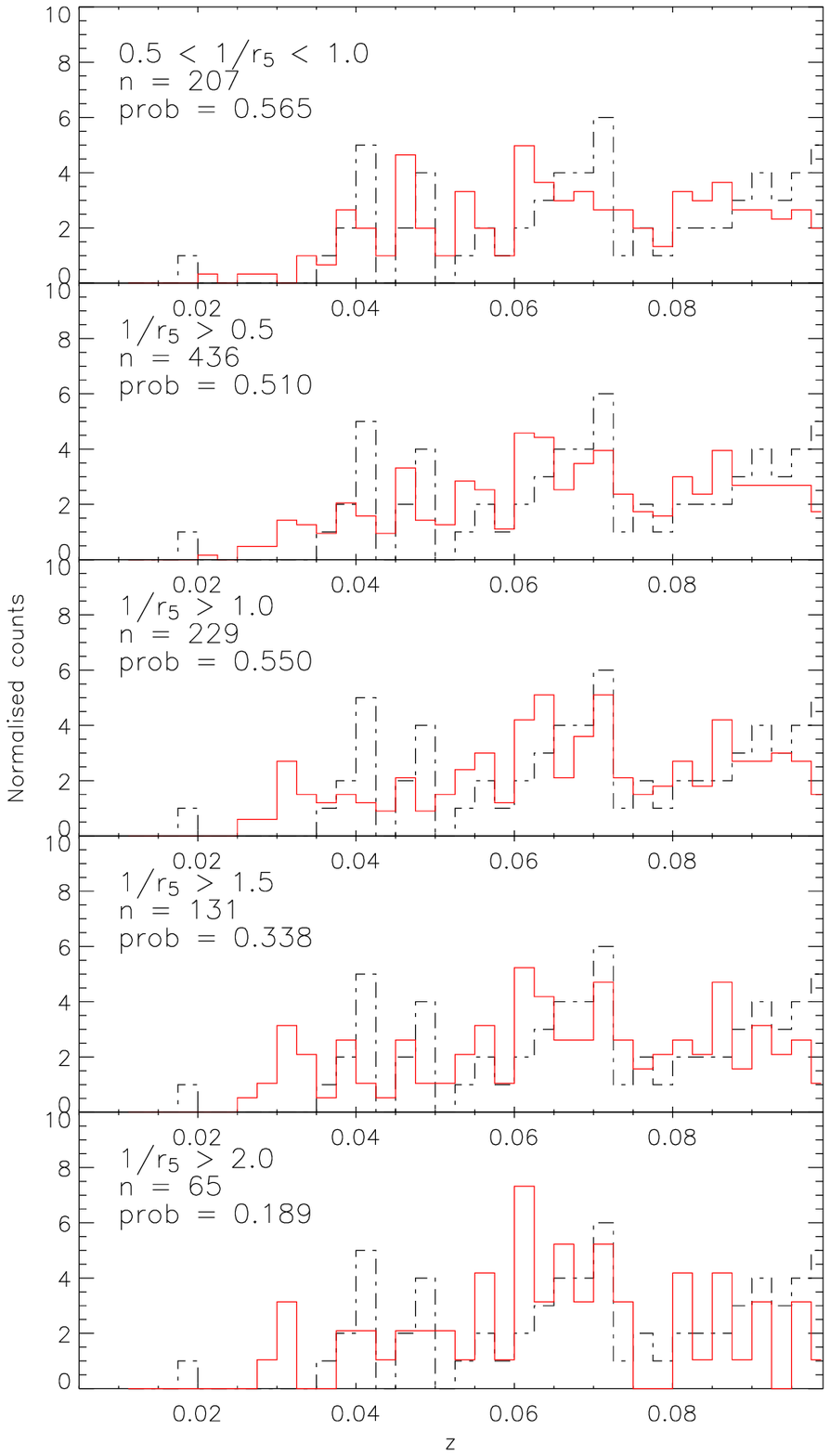}}
\end{tabular}}
\caption{Distribution of redshift for various density environments for 3 ranges of the velocity dispersion. The lines and labelling are as in fig.~\ref{fig:histograms}. {\bf Left:} $\sigma < 160\;\rm km\,s^{-1}$, {\bf middle:} $160 \leq \sigma \leq 230\;\rm km\,s^{-1}$, {\bf right:} $\sigma > 230\;\rm km\,s^{-1}$. The curves are normalised to the number of objects in the low density comparison sample which contains, 185, 473, and 68 objects respectively for the 3 ranges.}
\label{fig:sigbins}
\end{figure*}

In order to test whether these differences affect only the low mass objects or
are general to the early-type population as a whole (at least for objects
brighter than -20.45 in R-band) we have repeated our analysis for 3 bins in
velocity dispersion chosen to represent low, intermediate and high mass systems:
$\sigma < 160\;\rm km\,s^{-1}$, $160 \leq \sigma \leq 230\;\rm km\,s^{-1}$ and
$\sigma > 230\;\rm km\,s^{-1}$. The luminosity and redshift distributions of
these sub-samples are shown in Fig.~\ref{fig:sigbins}, where it can be seen that
they are similarly distributed. The results of the analysis for these 3 ranges of
velocity dispersion are shown in the rightmost 3 panels for each index of Fig.~\ref{fig:monte-together}.

Bearing in mind the larger errors on the results for the sample when divided by
velocity dispersion (especially for the low and high sigma sub-samples) we find
that most of the trends seen for the sample as a whole are also seen for the sub-sample 
with $\sigma < 160\;\rm km\,s^{-1}$. Indeed, the magnitude of the
environmental index differences is larger for these low-mass objects for all the
indices in which some effect is seen for the sample as a whole. The sub-sample with $\sigma > 230\;\rm km\,s^{-1}$ shows
effects with environment that are generally weaker than those for the low-mass sub-sample
and for the sample as a whole such that the magnitude of the environmental effect
falls from panel 2 to panel 4 of Fig.~\ref{fig:monte-together}. CN1 and Mg1 show
this trend especially clearly. However there are exceptions to this trend. G4300,
Hb, Fe5015 and Fe5270 seem to show stronger trends with environment for more
massive objects although none of the trends are as convincing as those in the
opposite sense.

\begin{figure*}
{\centering \begin{tabular}{ccc}
\resizebox*{0.33\textwidth}{!}{\includegraphics{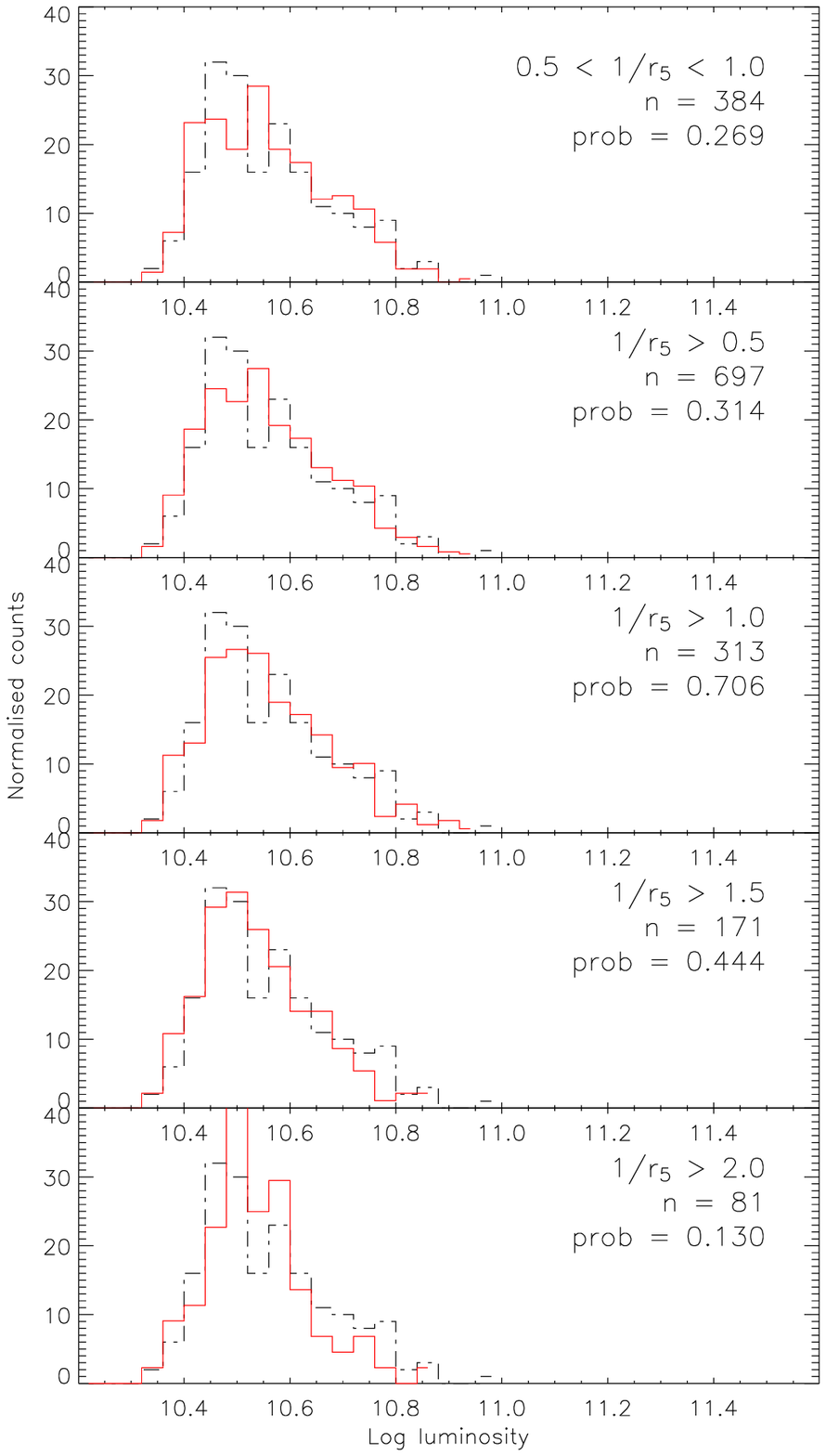}} &
\resizebox*{0.33\textwidth}{!}{\includegraphics{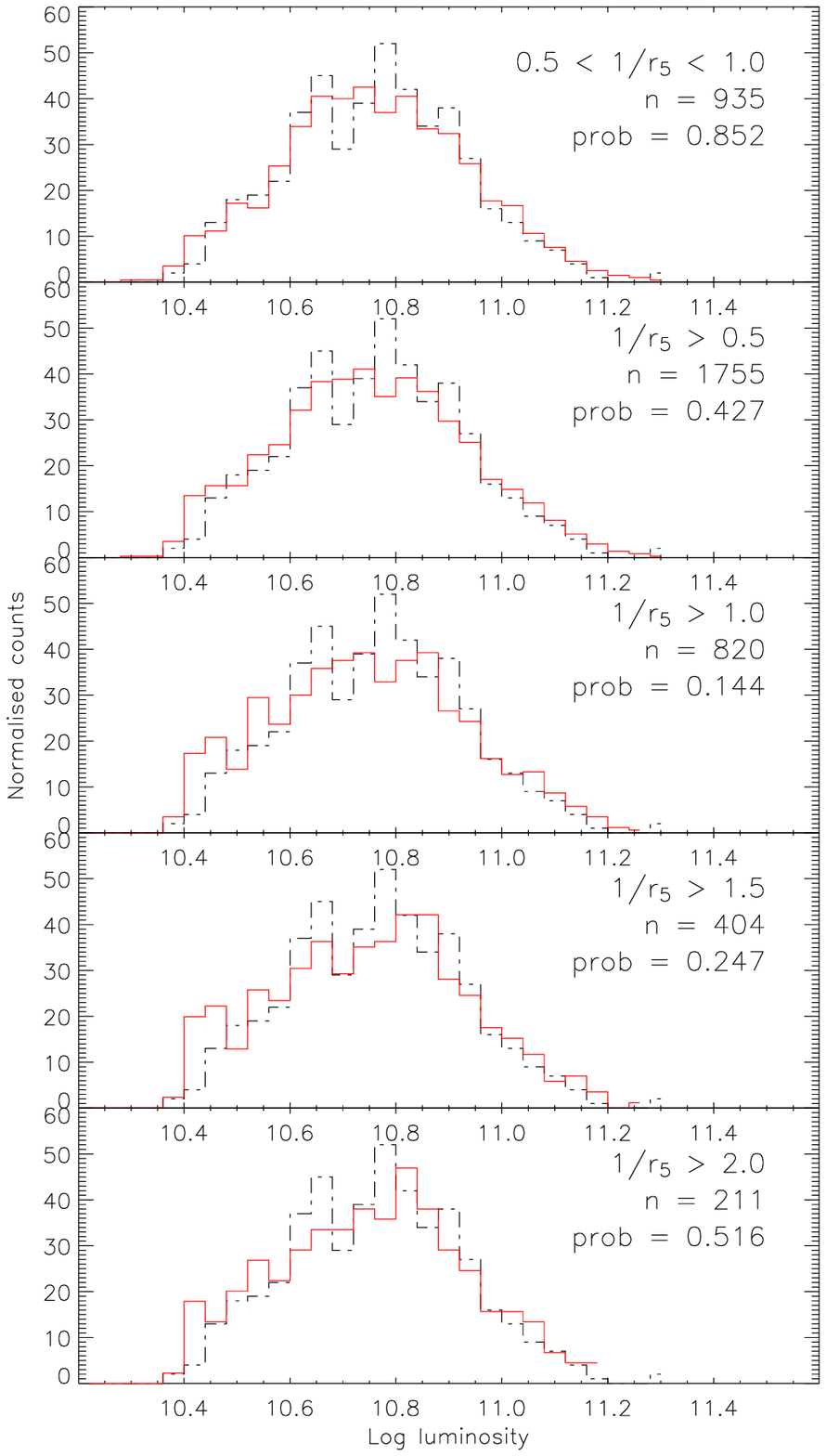}} &
\resizebox*{0.33\textwidth}{!}{\includegraphics{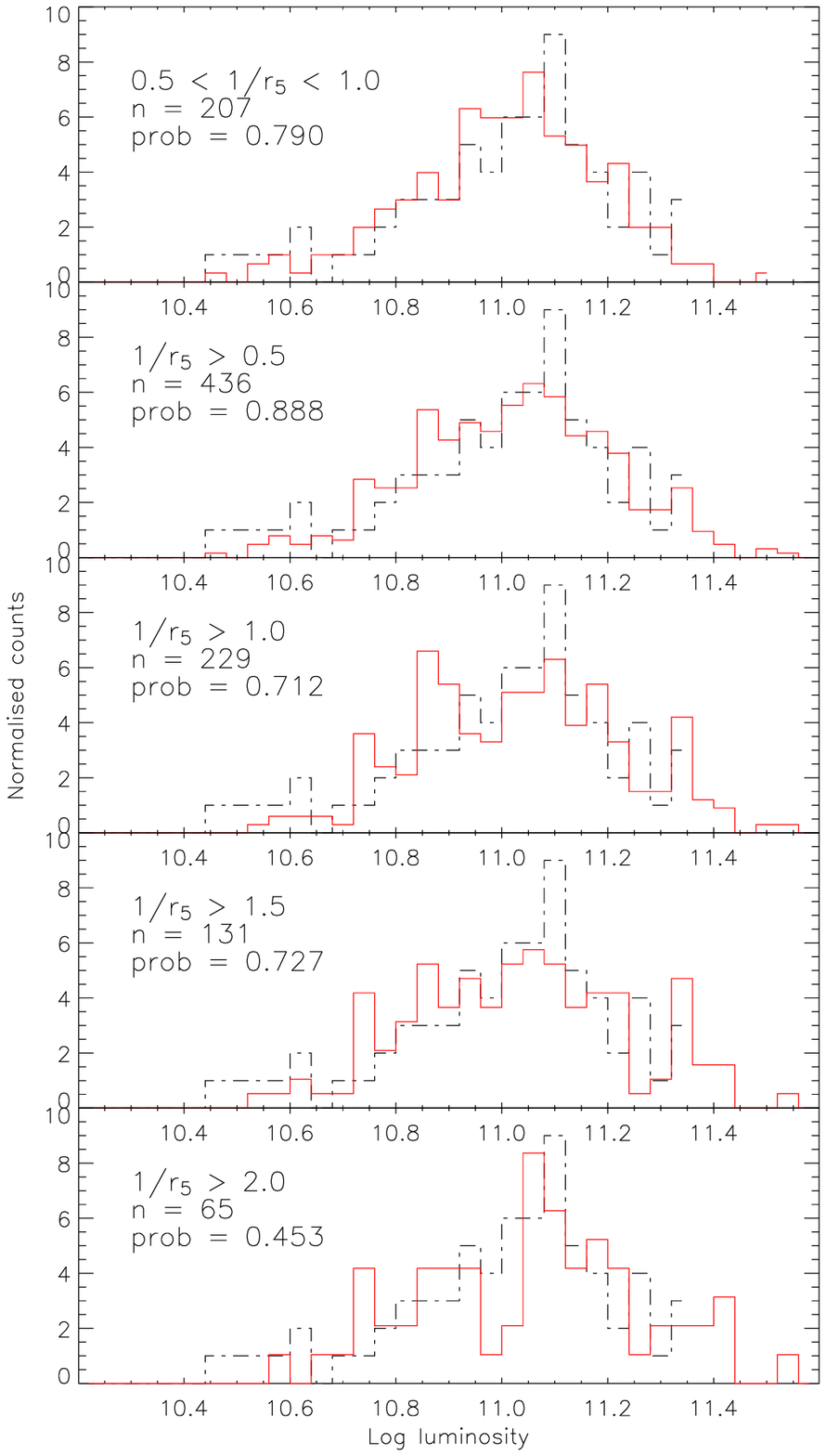}}
\end{tabular}}
\caption{As for fig.~\ref{fig:sigbins} for the distribution of luminosity.}
\label{fig:sigbins-lum}
\end{figure*}

\label{lastpage}
\end{document}